\documentclass[pra,twocolumn,superscriptaddress,showpacs,amsmath,amstex,amssymb,citeautoscript]{revtex4-2}

\usepackage{booktabs}
\usepackage{standalone}
\usepackage[T1]{fontenc}
\usepackage[utf8]{inputenc}
\usepackage{lipsum}
\usepackage{amsmath}
\usepackage{amssymb}
\usepackage{bbm}
\usepackage{braket}
\usepackage{xcolor}
\usepackage{pifont}
\usepackage{graphicx}
\usepackage[mathscr]{euscript}
\usepackage[shortlabels]{enumitem}
\usepackage[justification=justified]{subcaption}
\usepackage{tikz}
\usepackage{graphicx}
\usepackage{lipsum}
\allowdisplaybreaks
\usepackage{float}
\usepackage{graphicx}
\usepackage{dsfont}
\usepackage{comment}

\makeatletter
\let\stdtextcites\@textcitedefault
\let\@textcitedefault\textcite
\makeatother
\usepackage[colorlinks=true]{hyperref}  
\hypersetup{
    bookmarks=true,         % show bookmarks bar?
    unicode=false,          % non-Latin characters 
    pdftoolbar=true,        % show Acrobat
    pdfmenubar=true,        % show Acrobat 
    pdffitwindow=false,     % window fit to page when opened
    pdfstartview={FitH},    % fits the width of the page to the window
    pdftitle={Fractionalized Altermagnets},    % title
    pdfauthor={Sobral, Mandal, Scheurer},     % author
    pdfsubject={},   % subject of the document
    pdfcreator={},   % creator of the document
    pdfproducer={}, % producer of the document
    pdfkeywords={} {} {}, % list of keywords
    pdfnewwindow=true,      % links in new window
    colorlinks=true,       % false: boxed links; true: colored links
    linkcolor=blue, %red,          % color of internal links (change box color with linkbordercolor)
    citecolor=blue,        % color of links to bibliography
    filecolor=magenta,      % color of file links
    urlcolor=blue           % color of external links
}

\newcommand{\equref}[1]{Eq.~(\ref{#1})}

\newcommand{\secref}[1]{Sec.~\ref{#1}}
\newcommand{\figref}[1]{Fig.~\ref{#1}}
\newcommand{\refcite}[1]{Ref.~\onlinecite{#1}} 
\newcommand{\refscite}[1]{Refs.~\onlinecite{#1}}

\newcommand{\tableref}[1]{Table~\ref{#1}}
\newcommand{\appref}[1]{Appendix~\ref{#1}}

\newcommand{\pdagger}{{\phantom{\dagger}}}
\newcommand{\diff}{\mathrm{d}}

\renewcommand{\approx}{\simeq}

\renewcommand{\vec}[1]{\boldsymbol{#1}}
\newcommand{\rotatedsquare}[4]{
  \mathop{{\tikz[baseline=(current bounding box.center), scale=0.7] {
    \node[inner sep=0, minimum size=1em] (s) {\rotatebox[origin=c]{45}{$\square$}};
    \node[font=\scriptsize] at ([shift={(-1em,0em)}]s.center) {$#1$};
    \node[font=\scriptsize] at ([shift={(0em,1.2em)}]s.center) {$#2$};
    \node[font=\scriptsize] at ([shift={(1em,0em)}]s.center) {$#3$};
    \node[font=\scriptsize] at ([shift={(0em,-1.2em)}]s.center) {$#4$};
    }}}
}

\definecolor{wrongultramarine}{rgb}{1,0.5,0}

\begin{document}

\title{Fractionalized Altermagnets: \\ from neighboring and altermagnetic spin-liquids to spin-symmetric band splitting}

\author{João Augusto Sobral}
\affiliation{Institute for Theoretical Physics III, University of Stuttgart, 70550 Stuttgart, Germany}

\author{Subrata Mandal}
\affiliation{Institute for Theoretical Physics III, University of Stuttgart, 70550 Stuttgart, Germany}

\author{Mathias S.~Scheurer}
\affiliation{Institute for Theoretical Physics III, University of Stuttgart, 70550 Stuttgart, Germany}

\begin{abstract}
We study quantum-fluctuation-driven fractionalized phases in the vicinity of altermagnetic order. First, the long-range magnetic orders in the vicinity of collinear altermagnetism are identified; these feature a non-coplanar ``orbital altermagnet'' which has altermagnetic symmetries in spin-rotation invariant observables. We then describe neighboring fractionalized phases with topological order reached when quantum fluctuations destroy long-range spin order, within Schwinger-boson theory and an SU(2) gauge theory of fluctuating magnetism. Discrete symmetries remain broken in some of the fractionalized phases, with the orbital altermagnet becoming an “altermagnetic spin liquid”. We compute the electronic spectral function in the doped system, which is characterized by split Fermi surfaces with preserved spin-rotation symmetry. 
\end{abstract}

%\date{\today}
\maketitle
%\tableofcontents

\section{Introduction}
In the last few years, there has been an increasing amount of research on ``altermagnets'' (AMs)
\cite{yuan_giant_2020,smejkal_beyond_2022,smejkal_emerging_2022,ahn_antiferromagnetism_2019,bhowal_magnetic_2022,cuono_orbital-selective_2023,guo_spin-split_2023,maier_weak-coupling_2023,mazin_altermagnetism_2023,oganesyan_quantum_2001,sato_altermagnetic_2023,steward_dynamic_2023,turek_altermagnetism_2022,ouassou_dc_2023,mazin_prediction_21,hayami_momentum_19,smejkal_crystal_20,brekke_two-dimensional_2023,das_realizing_2023,fakhredine_interplay_2023,fang_quantum_2023,leeb_spontaneous_2023,mazin_induced_2023,smejkal_chiral_2023,sun_spin_2023,gao_ai-accelerated_2023,beenakker_phase-shifted_2023,thermal_transport,RafaelsPaper,andreev,zhang2023finitemomentum,brekke_two-dimensional_2023,giil2023superconductoraltermagnet,majorana1,majorana2,majorana3,orientation_altermagnet,antonenko2024mirror,yu2024altermagnetism,wei2023gapless,mcclarty2023landau,banerjeeAltermagneticSuperconductingDiode2024,yershovFluctuationInducedPiezomagnetism2024,yershovFluctuationInducedPiezomagnetism2024,PhysRevB.108.224421,PhysRevB.108.L180401,2024arXiv240218629C,2024arXiv240910034J,2024arXiv240910712D,2024arXiv240701513S,2023arXiv230914812X,2024arXiv240317050B,2024arXiv240715836O,2024arXiv240215616R, zhaokondo2024,aoyama_piezomagnetic_2023,bai_efficient_2023,krempasky_altermagnetic_2023,lee_broken_2023,reimers_direct_2023,feng_anomalous_2022,bose_tilted_2022,2024arXiv240913504L, Ma2021May, ferrariAltermagnetismShastrySutherlandLattice2024, Solovyev1997Apr, Noda2016May, Naka2019Sep, Naka2021Mar, Okugawa2018Jan}. Unlike antiferromagnets, the total magnetic moment in AMs does not vanish as a result of translations combined with time-reversal ($\Theta$) symmetry but, instead, due to the product $\Theta g$ where $g$ is a point-group symmetry, e.g., four-fold rotational symmetry $g=C_{4z}$ on the square lattice \cite{smejkal_beyond_2022}.
While many of the above mentioned studies deal with the non-trivial impact of altermagnetic moments on quantum physics, e.g., on the electronic band structure or on the properties of superconductivity, studying the impact of quantum fluctuations on the altermagnetic order itself is far less explored. In this work, we address the regime of strong quantum fluctuations which restore spin-rotation (SR) invariance in AMs and in other neighboring non-trivial magnetic orders, both in insulators and metallic phases. This is also relevant in light of recent experiments that indicate the interplay between altermagnetism and frustration for thin films of Mn$_5$Si$_3$ \cite{Biniskos2022Mar, Surgers2024Aug}  and CsCr$_3$Sb \cite{2023arXiv230914812X}.

More specifically, we start from  magnetically ordered phases close to a square-lattice AM in a checkerboard Heisenberg model \cite{earliestCheckerboard,bernierPlanarPyrochloreAntiferromagnet2004,bishopFrustratedHeisenbergAntiferromagnet2012,sadrzadehPhaseDiagramJ1J22015,yershovFluctuationInducedPiezomagnetism2024,2024arXiv240218629C,Li_2015, Fouet2003Feb, Canals2002Apr, Tchernyshyov2003Oct, Chan2011Dec} supplemented by ring-exchange terms. Apart from the collinear AM and other non-collinear states that preserve all symmetries modulo SRs (i.e., will appear symmetric in SR-invariant observables), there are also nematic spin orders and an orbital AM. The last phase, which is connected to the collinear AM by a second-order phase transition in our model, is characterized by scalar spin chiralities which are staggered on the lattice in a way that preserves $\Theta C_{4z}$ and translation, akin to the spin order of the prototypical AM. We characterize the associated spin-liquid phases obtained by ``quantum melting'' any of these magnetic orders. We present Schwinger-Boson (SB) \cite{auerbach2012interacting,auerbach2010schwinger,PhysRevLett.66.1773,bernierPlanarPyrochloreAntiferromagnet2004,yangSchwingerBosonSpinliquid2016,samajdarThermalHallEffect2019} ans\"atze for each of them and demonstrate that these spin liquids inherit discrete broken symmetries. Most notably, the orbital AM becomes an altermagnetic spin-liquid which is characterized by the same altermagnetic spatial arrangement of scalar spin chiralities. Finally, we supplement this with an SU(2) gauge theory \cite{PhysRevLett.61.467,PhysRevLett.65.2462,SchriefferRotatingRefFrame,PhysRevB.80.155129,PhysRevB.81.115129,chatterjeeIntertwiningTopologicalOrder2017,OurPNAS,PhysRevB.98.235126}, which allows us to generalize these fractionalized orders to the doped system. Remarkably, the spectral function of the itinerant electrons exhibits peaks similar to the spin-splitting in AMs, while maintaining SR invariance.

%==============================================
\vspace{0.5em}
\section{Classical phase diagram}
\begin{figure}[tb]
   \centering
   \includegraphics[width=\linewidth]{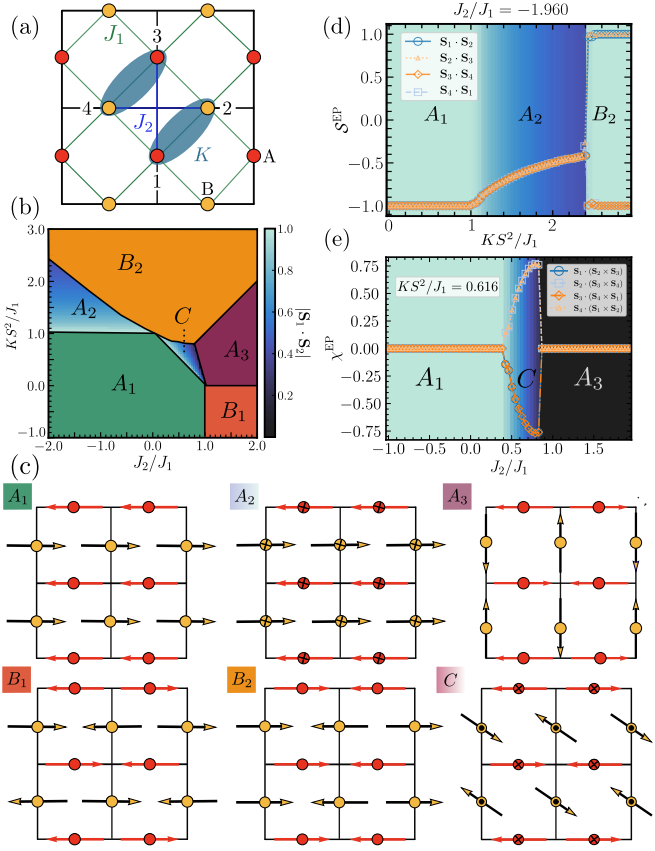}%{panela2.pdf}
   \caption{(a) Bravais lattice (black squares), spins (red, orange sublattices), and couplings $J_{1}$, $J_{2}$ and $K$ of the model \eqref{eq:heisenbergmodel}. (b) Classical phase diagram, with the corresponding phases defined in (c) and \tableref{table:classicalphases}. (d) Nearest-neighbor spin products $\vec{\mathcal{S}}^{\text{EP}}$ and (e) scalar spin chiralities $\vec{\chi}^{\text{EP}}$ along two different one-dimensional cuts through (b). The background coloring  in (d-e) refers to $|\vec{S}_{1}\cdot\vec{S}_{2}|$ as defined in the colorbar in (b).}
   \label{fig:classicalPD}
\end{figure}

Although the concepts are more general, we start, for concreteness, our discussion with a minimal Heisenberg model. To be able to capture AMs, we need two sublattices ($A$, $B$) per unit cell, which we place on the bonds of a square lattice, see \figref{fig:classicalPD}(a). Denoting the spin operators on bond $i$ by $\hat{\vec{S}}_i$, the Hamiltonian reads as
\begin{align}
&\mathcal{H} = J_{1}\sum_{\langle i, j \rangle}  \hat{\boldsymbol{S}}_i \cdot \hat{\boldsymbol{S}}_j  
+ J_{2}\sum_{\langle \langle i, j  \rangle\rangle}  \hat{\boldsymbol{S}}_i \cdot \hat{\boldsymbol{S}}_j \label{eq:heisenbergmodel} \\
&\,\, + K \hspace{-0.5em} \sum_{\rotatedsquare{j}{k}{l}{i}}\left[\left(\hat{\boldsymbol{S}}_i \cdot \hat{\boldsymbol{S}}_j\right)\left(\hat{\boldsymbol{S}}_k \cdot \hat{\boldsymbol{S}}_{l}\right)+\left(\hat{\boldsymbol{S}}_i \cdot \hat{\boldsymbol{S}}_{l}\right)\left(\hat{\boldsymbol{S}}_k \cdot \hat{\boldsymbol{S}}_j\right)\right].  \nonumber
\end{align}
If we only included the first term---the nearest-neighbor exchange interaction, which we choose to obey $J_1 > 0$ here---the model would become identical to a square-lattice antiferromagnet (with the lattice rotated by $45^\circ$). Adding the next-nearest neighbor couplings $ J_2$ (solid blue lines in \figref{fig:classicalPD}a) leads to the checkerboard model \cite{earliestCheckerboard,bernierPlanarPyrochloreAntiferromagnet2004,bishopFrustratedHeisenbergAntiferromagnet2012,sadrzadehPhaseDiagramJ1J22015,yershovFluctuationInducedPiezomagnetism2024,2024arXiv240218629C,Li_2015, Fouet2003Feb, Canals2002Apr, Tchernyshyov2003Oct, Chan2011Dec}. To lift degeneracies and stabilize additional phases, we further include the ring exchange term in the second line of \equref{eq:heisenbergmodel}; it couples any four spins, $i,j,k,l$, on bonds with a common vertex in a way that preserves the $C_{4v}$ point group symmetries, time-reversal $\Theta$, SR, as well as Bravais-lattice translational symmetry. 

We first analyze the magnetic phases in the classical limit and make the ansatz 
\begin{equation}
    \vec{S}_{\ell}(\vec{R}) \propto \hat{\vec{a}} \cos ( \vec{Q}\cdot \vec{R} -\varphi_\ell) + \hat{\vec{b}} \sin ( \vec{Q} \cdot \vec{R} -\varphi_\ell) + \hat{\vec{c}} \, \eta_\ell,
    \label{eq:ansatzeclassical}
\end{equation}
for the direction of the magnetization on sublattice $\ell =A,B$ in square-lattice unit cell $\vec{R}$; here $\hat{\vec{a}}$, $\hat{\vec{b}}$, $\hat{\vec{c}}$ are three arbitrary three-component orthonormal vectors. Without loss of generality, we set $\varphi_A=0$ and $\varphi_B \equiv \varphi$ and minimize the classical energy with respect to the variational parameters $\{Q_{x}, Q_{y},\varphi, \eta_{A}, \eta_{B}\}$ (see \appref{app:classicalpd}).

In agreement with the nearest-neighbor square-lattice antiferromagnet model, the system favors an antiparallel alignment of the spins on the two sublattices, $\vec{S}_A(\vec{R}) = -\vec{S}_B(\vec{R}) =\hat{\vec{a}}$, for small $K$ and $J_2$, see phase A$_1$ in \figref{fig:classicalPD}(b,c). Importantly, though, finite $K$ and $J_2$ in \equref{eq:heisenbergmodel} break the artificial translational symmetry at $J_2=K=0$ relating the two sublattices $A$ and $B$; instead, they are related by a $C_{4z}$ rotational symmetry in the full model, making A$_1$ a collinear AM. It has vanishing total magnetization, $\vec{M} = \sum_{\vec{R},\ell} \vec{S}_{\ell}(\vec{R}) = 0$, as guaranteed by $C_{4z}\Theta$. Modulo global SRs, all magnetic point-group symmetries are preserved, which will also play an important role to our later discussion of nearby spin liquids. Since this implies that all SR-invariant observables must also respect all symmetries of our model, we can conveniently probe these symmetries using the nearest neighbor scalar and (anti-clockwise) triple products $\vec{\mathcal{S}}^{\text{EP}} =(\vec{S}_{1}\cdot \vec{S}_{2},\vec{S}_{2}\cdot \vec{S}_{3},\vec{S}_{3}\cdot \vec{S}_{4},\vec{S}_{4}\cdot \vec{S}_{1})^{T} $ and $\vec{\chi}^{\text{EP}}=\left(\vec{S}_{1}\cdot \left(\vec{S}_{2}\times \vec{S}_{3}\right), \right.$ $\left.\vec{S}_{2}\cdot\left(\vec{S}_{3}\times\vec{S}_{4}\right), \right.$ $\left.\vec{S}_{3}\cdot\left(\vec{S}_{4}\times \vec{S}_{1}\right), \vec{S}_{4}\cdot \left(\vec{S}_{1}\times \vec{S}_{2}\right)\right)^{T}$, respectively, where the sites are labeled as in \figref{fig:classicalPD}(a). Indeed, we can see from \tableref{table:classicalphases} that all components of $\vec{\mathcal{S}}^{\text{EP}}$ are identical (isotropic/non-nematic) and $\vec{\chi}^{\text{EP}} = 0$ (preserved $\Theta$/non-chiral). 

\begin{center}
\begin{table*}[tb]
\captionof{table}{Summary of the magnetic phases for the spin model (\ref{eq:heisenbergmodel}). We list one possible set of parameters for \equref{eq:ansatzeclassical} and the values of $\vec{\mathcal{S}}^{\text{EP}}$, $\vec{\chi}^{\text{EP}}$ ($s_0$, $s_0'$, $\xi_0 \in \mathbb{R}^{+}$ are non-universal parameters depending on the couplings). 
We further indicate whether the total magnetization $\vec{M}$ is finite and the magnetic point symmetries of spin-rotation-invariant observables (through their generators); these are also the symmetries of the associated spin liquids in \figref{fig:SchwingerBoson} and the itinerant fractionalized phases with topological order discussed below. The invariant gauge group (IGG) of these two descriptions coincide and are listed in the last column.}
\label{table:classicalphases}
\begin{ruledtabular}
\begin{tabular}{cccccccc}
Label & Type & $\vec{\mathcal{S}}^{\text{EP}}$ & $\vec{\chi}^{\text{EP}}$ & $(Q_{x}, Q_{y}, \varphi, \eta_{A}, \eta_{B})$ & $\vec{M}$  & Symmetries & IGG  \\ \hline
$A_1$ & collinear AM (isotropic, non-chiral)    & $-\left(1,1,1,1\right)$ & $\left(0,0,0,0\right)$ & $\left(0,0, \pi, 0, 0\right)$ & $0$  & $C_{4z}, \sigma_v, \Theta$  & U(1)   \\
\midrule
$A_2$ & canted AM (isotropic, non-chiral)    & $-\left(s_{0},s_{0},s_{0},s_{0}\right)$ & $\left(0,0,0,0\right)$ & $\left(0, 0 ,\pi,\eta,\eta\right)$ & $\neq 0$ & $C_{4z}, \sigma_v, \Theta$ & $\mathbb{Z}_2$  \\
\midrule
$A_3$ & orthogonal AFM (isotropic, non-chiral)    & $ \left(0, 0, 0, 0\right)$ & $\left(0,0,0,0\right)$ & $\left(\pi,\pi,\pi/2, 0, 0 \right)$ & $0$ & $C_{4z}, \sigma_v, \Theta$ & $\mathbb{Z}_2$ \\
\midrule
$B_1$ &  collinear nematic (non-chiral)  & $\left(-1,1,-1,1\right)$ & $\left(0,0,0,0\right)$ & $\left(\pi,\pi,0, 0, 0 \right)$ & $0$ & $C_{2z}, \sigma_d$ & U(1)  \\
\midrule
$B_2$ & collinear odd parity (non-chiral)   & $ \left(1, 1, -1, -1\right)$ & $\left(0,0,0,0\right)$ & $\left(\pi,0,0, 0, 0 \right)$ & $ 0$ & $\sigma_v, \Theta$ & U(1) 

\\
\midrule
$C$ & orbital AM (isotropic, staggered chirality)    & $-\left(s'_{0},s'_{0},s'_{0},s'_{0}\right)$ & $\left(-\xi_0,\xi_0,-\xi_0,\xi_0\right)$ & $\left(\pi , \pi,\pi/2,\eta, -\eta \right)$ & 0 & $C_{4z}\Theta, \sigma_v\Theta$ & $\mathbb{Z}_2$  \\
\end{tabular}
\end{ruledtabular}
\end{table*}
\end{center}

Turning on sufficiently large $K > 0$, it becomes energetically favorable to reduce the magnitude of the nearest-neighbor spin products, by developing a finite canting. This leads to the second-order phase transition into the canted AM ($A_2$), $\vec{S}_{A,B}(\vec{R}) = \pm \hat{\vec{a}} + \eta \hat{\vec{c}}$, see \figref{fig:classicalPD}(b-d). As summarized in \tableref{table:classicalphases}, this leads to a finite magnetization $\vec{M} \neq 0$ while the symmetries (modulo SR) are identical to those of A$_1$. In the phase diagram in \figref{fig:classicalPD}(b), there is one more fully symmetric phase, denoted by A$_3$; the sufficiently large $J_2$ leads to finite $Q_{x,y} = \pi$ while positive $K$ favors perpendicular spins in the two sublattices ($\varphi=\pi/2$), see \figref{fig:classicalPD}(c). As a result of translational symmetry the magnetization vanishes, and the state can be thought of as an antiferromagnet with four spins in the enlarged unit cell. 

Furthermore, the phase diagram contains two nematic phases, denoted by B$_{1,2}$ which break $C_{4z}$ rotational symmetry in SR-invariant observables while retaining $\Theta$ symmetry, see \figref{fig:classicalPD}(c) and \tableref{table:classicalphases}. The phase B$_1$ is reached via a first-order transition from A$_3$ by reversing the sign of $K$ leading to parallel spins in the unit cell ($\varphi=0$). Meanwhile B$_2$ is stabilized when $K>0$ dominates the energetics leading to a phase with only one of $Q_x$, $Q_y$ non-zero (and equal to $\pi$).

Finally, in the most frustrated region of the phase diagram in \figref{fig:classicalPD}(b), a non-coplanar phase (C) is stabilized. As can be seen in \figref{fig:classicalPD}(e), this state is reached from the collinear AM A$_1$ via a second-order phase transition, where the scalar spin chiralities become non-zero, breaking $\Theta$ in SR-invariant observables. Importantly, the chiralities are staggered, $\vec{\chi}^{\text{EP}} \propto (-1,1,-1,1)$, in a way that preserves $C_{4z}\Theta$ (and $\sigma_v \Theta$). As such, phase C behaves like an AM in SR-invariant observables, i.e., is associated with (when adding charge carriers, see \secref{Itinerant} below) a special arrangement of orbital currents \cite{PhysRevLett.89.247003,PhysRevB.55.14554,chatterjeeIntertwiningTopologicalOrder2017,PhysRevB.98.235126}, where the corresponding orbital magnetic moments are finite locally but vanish globally --- not as a result of translational symmetry but due to $C_{4z}\Theta$. For this reason, we refer to phase C as an ``orbital AM'' \cite{yu2024altermagnetism}.

%==============================================
\vspace{0.5em}
\section{Nearby spin liquids} Having established the classical phase diagram, we next discuss spin-liquid phases that can be reached from the respective magnetically ordered phases once quantum fluctuations restore SR invariance. To this end, we will first use a SB description \cite{auerbach2012interacting,auerbach2010schwinger} where the spin operators in \equref{eq:heisenbergmodel} are rewritten as $\hat{\vec{S}}_{i}= \hat{b}^{\dagger}_{i \sigma} {\vec{\sigma}}_{\sigma \sigma '} \hat{b}^\pdagger_{i \sigma'}/2$, introducing a local U(1) gauge redundancy, $\hat{b}_{j,\sigma} \rightarrow e^{i\phi_j }\hat{b}_{j,\sigma}$; here $\vec{\sigma}$ are Pauli matrices and $\hat{b}_{i,\sigma}$ canonical bosons. Since this enlarges the Hilbert space of spin-$S$ operators, the local constraint $\hat{n}_i = 2S$, $\hat{n}_i = \hat{b}^{\dagger}_{i \sigma} \hat{b}^\pdagger_{i \sigma}$, is imposed. Decoupling the resulting boson-boson interactions in the spin Hamiltonian, leads to a free boson Hamiltonian
\begin{equation}
\mathcal{H}_{b} =  \dfrac{1}{2} \sum_{i,j}  (\mathcal{B}^{\ast}_{ij} \hat{b}_{i \sigma} \hat{b}^{\dagger}_{j \sigma} - \mathcal{A}^{\ast}_{ij} \sigma \hat{b}_{i \sigma}  \hat{b}_{j - \sigma}  + \text{H.c.}) - \mu \sum_{i} \hat{n}_i, \nonumber
\end{equation}
where $\mathcal{A}_{ij}, \mathcal{B}_{ij} \in \mathbb{C}$ can be computed within self-consistent mean-field theory; the same applies for the Lagrange multiplier $\mu$ introduced to treat the local constraint on average. 
However, as this mean-field approach is typically not quantitatively reliable, we here take a different perspective: we systematically study possible \textit{ans\"atze} for $\mathcal{A}_{ij}, \mathcal{B}_{ij}$, starting with only nearest-neighbor terms and adding further-neighbor couplings until we find a spin-liquid where Bose condensation (see \appref{app:SchwingerBoson}). leads to a (apart from global SRs) unique magnetic ground state that is identical to one of the phases in \figref{fig:classicalPD}(b) and \tableref{table:classicalphases}; in this way, we can associate a spin-liquid with each of these magnetic phases.

\begin{figure}[htp!]
   \centering
    \includegraphics[width=\linewidth]{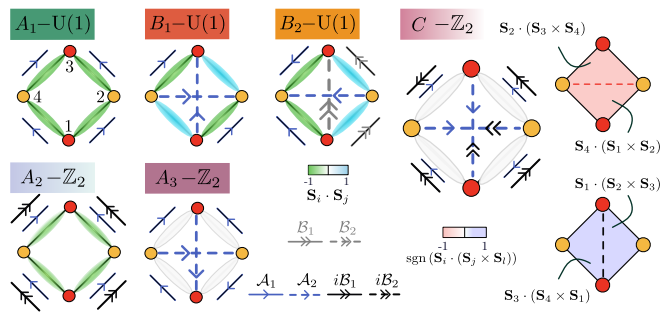}
    \caption{Diagrammatic definition of the Schwinger-Boson ans\"atze, where blue (black and grey) arrows indicate real $\mathcal{A}_{ij}$ (imaginary and real $\mathcal{B}_{ij}$) on the respective bonds, for the spin-liquids associated with the magnetic phases in \figref{fig:classicalPD}(b,c). We further indicate the expectation values of $\vec{\mathcal{S}}^{\text{EP}}$ using colored ellipses and, for phase C, of $\vec{\chi}^{\text{EP}}$ (only showing two triple products in each of the nearest two unit cells for clarity) in the respective spin liquids. The patterns extend through the entire lattice.
    } \label{fig:SchwingerBoson}
\end{figure}

For all of them, restricting the model to nearest and next-nearest-neighbor bonds $i,j$ in $\mathcal{A}_{ij}, \mathcal{B}_{ij}$ which (upon proper gauge choice) do not increase the unit cell turns out to be sufficient. We define all of these ans\"atze diagrammatically in \figref{fig:SchwingerBoson}. For instance, we see that the ansatz for A$_1$ only involves $\mathcal{A}_{ij}$ on nearest-neighbor bonds. Being left invariant by a U(1) gauge transformation consisting of phases of alternating signs in the two sublattices, it is a U(1) spin liquid. It is straightforward to show that the ansatz is invariant (modulo gauge transformations) under all symmetries of the model (\ref{eq:heisenbergmodel}), which we also demonstrated by computing the expectation values of $\vec{\mathcal{S}}^{\text{EP}}$ and $\vec{\chi}^{\text{EP}}$ in the spin-liquid phase: indeed, as also indicated in \figref{fig:SchwingerBoson}, we get $\braket{\vec{\mathcal{S}}^{\text{EP}}_l} = \textit{const}$ and $\braket{\vec{\chi}^{\text{EP}}} = 0$. 

The ansatz for A$_2$ has additional imaginary $\mathcal{B}_{ij}$ on nearest neighbor bonds, which not only control the canting in the associated magnetically ordered phase but also lead to a $\mathbb{Z}_2$ spin liquid. Just as the ansatz for A$_1$, this $\mathbb{Z}_2$ spin liquid also respects all symmetries of the Hamiltonian. Also for the third magnetic phase, A$_3$, which respects all symmetries modulo SRs, the associated SB ansatz, now with second-nearest-neighbor terms, see \figref{fig:SchwingerBoson}, leads to a fully symmetric $\mathbb{Z}_2$ spin liquid. 

In contrast, the spin liquids associated with the remaining phases have reduced symmetries. As one would expect, the spin liquids of the nematic magnetic phases B$_{1,2}$ are nematic themselves: as can be seen from the correlators $\braket{\vec{\mathcal{S}}^{\text{EP}}}$ shown along with the ans\"atze in \figref{fig:SchwingerBoson} and upon noting that still $\braket{\vec{\chi}^{\text{EP}}} = 0$ ($\Theta$ is not broken), they obey the exact same symmetries as listed in \tableref{table:classicalphases} for B$_{1,2}$. 
Finally, the spin liquid associated with the orbital AM also has reduced symmetries: we find isotropic correlators $\braket{\vec{\mathcal{S}}^{\text{EP}}_l} = \textit{const}$ but staggered scalar spin chiralities $\braket{\vec{\chi}^{\text{EP}}}\propto (-1,1,-1,1)$; while SR invariance is restored, the altermagnetic arrangement of orbital moments (being translation invariant but odd under $C_{4z}$) survive in the neighboring spin-liquid phase, which can thus be thought of as an ``altermagnetic spin liquid''. This phase hosts vison excitations associated with local distortions of the magnetization pattern \cite{exact2025} that could be detected with local experimental probes \cite{Jahin2024Jul}.

%==============================================
\section{Fractionalized itinerant systems}\label{Itinerant}
\begin{figure}[tb]
   \centering
    \includegraphics[width=\linewidth]{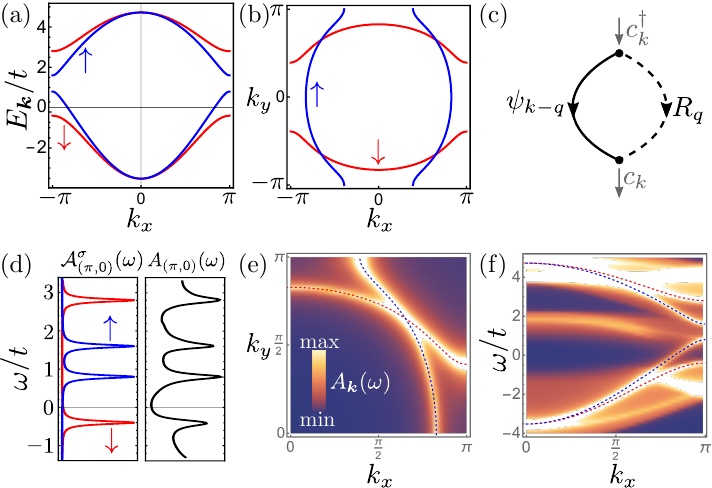}
    \caption{Band structure (a) along the $k_y=0$ line and (b) Fermi surfaces for the long-range A$_1$ order. The electronic Green's function is a convolution (c) of the chargon and spinon Green's function. (d) Electronic spectral function in the presence of long-range order ($\mathcal{A}^\sigma_{\vec{k}}(\omega)$, left) and in the fractionalized phase with spin-rotation invariance and topological order ($A_{\vec{k}}(\omega)$, right) at $\vec{k}=(\pi,0)$. In (e) and (f), the spectral function in the fractionalized phase is shown at $\omega = 0$ as a function of momentum $\vec{k}$ and as a function of energy $\omega$ and $k_x$ at $k_y=0$, respectively. We use $t_2/t_1=0.3$, chemical potential $\mu/t_1=-1.2$, $H_0/t_1 =1$, temperature $T/t_1=0.1$, and a life-time broadening of $\eta/t_1=0.05$.}
    \label{fig:ElectronicSpectra}
\end{figure}

One of the central aspects of AMs is their impact on the spectrum of itinerant electrons, creating a momentum-dependent spin splitting of their bands. We here address the related question but in the regime where the spin degrees of freedom are in a spin-liquid state. To this end, we assume that mobile electrons are added to the model studied above which we describe by the creation operators $\hat{c}^\dagger_{\vec{R},\ell,\sigma}$ of spin-$\sigma$ electrons on the bond $(\vec{R},\ell)$ in \figref{fig:classicalPD}(a). We include nearest ($t_1$) and next-nearest-neighbor ($t_2$) hopping \cite{das_realizing_2023} and couple the electrons locally to the magnetic moments $\vec{\Phi}_{\vec{R},\ell}$ as
\begin{equation}
    \mathcal{S}_c = \int \diff \tau \sum_{\vec{R},\ell} c^\dagger_{\vec{R},\ell,\sigma} \vec{\sigma}_{\sigma,\sigma'} c^\pdagger_{\vec{R},\ell,\sigma'} \cdot \vec{\Phi}_{\vec{R},\ell} \label{PhiCCoupling}
\end{equation}
where we switched to a path integral description with imaginary time $\tau$ (see \appref{ElectronicSpcFun} for more details). Assuming non-fluctuating long-range order, $\vec{\Phi}_{\vec{R},\ell} \rightarrow \vec{S}_{\ell}(\vec{R})$, and taking, e.g., the collinear AM order $\vec{S}_{A,B}(\vec{R}) = \pm \hat{\vec{e}}_z$ of phase A$_1$, we obtain the bands and Fermi surface with the characteristic spin splitting of AMs shown in \figref{fig:ElectronicSpectra}(a,b).   

To describe the regime of strong quantum fluctuations, where SR invariance is restored, we employ the approach of \refscite{PhysRevLett.61.467,PhysRevLett.65.2462,SchriefferRotatingRefFrame,PhysRevB.80.155129,PhysRevB.81.115129,chatterjeeIntertwiningTopologicalOrder2017,OurPNAS,PhysRevB.98.235126}, where a transformation into a rotating reference frame in spin space is performed using the dynamical bosonic SU(2) matrices $R(\tau)$ in 
$c_{\vec{R},\ell,\sigma}(\tau) = [R_{\vec{R},\ell}(\tau)]_{\sigma,\alpha} \psi_{\vec{R},\ell,\alpha}(\tau)$. 
Physically, $R$ capture the spin degrees of freedom and are directly related to the bosons in our SB approach above, while the charge is carried by the fermionic ``chargons'' $\psi$. This rewriting introduces a local SU(2) redundancy as the physical degrees of freedom are unchanged under $R_{\vec{R},\ell} \rightarrow R_{\vec{R},\ell} V^\dagger_{\vec{R},\ell}$ and $\psi_{\vec{R},\ell,\alpha} \rightarrow V_{\vec{R},\ell}\psi_{\vec{R},\ell}$ for arbitrary SU(2) matrices $V_{\vec{R},\ell}$. We introduce the field $\vec{H}_{\vec{R},\ell}$ via
\begin{equation}
    \vec{H}_{\vec{R},\ell} \cdot \vec{\sigma} = \vec{\Phi}_{\vec{R},\ell} \cdot R^\dagger_{\vec{R},\ell} \vec{\sigma} R^\pdagger_{\vec{R},\ell},
\end{equation}
which can be intuitively thought of as $\vec{\Phi}$ represented in a rotating reference frame; in the emergent gauge theory, $\vec{H}_{\vec{R},\ell}$ plays the role of the Higgs field.
By construction, \equref{PhiCCoupling} can be compactly restated as a coupling between the chargons and the Higgs field [see \equref{HiggsFieldCoupling}].

If both $R_{\vec{R},\ell}$ and $\vec{H}_{\vec{R},\ell}$ condense, we will just reobtain long-range magnetic order; upon choosing a gauge where $\braket{R_{\vec{R},\ell}} = \mathbbm{1}$, the magnetic texture is equal to the texture of $\braket{\vec{H}_{\vec{R},\ell}}$, allowing us to connect the Higgs-field configurations to the classical magnetic orders studied above. When the bosons are gapped, $\braket{R_{\vec{R},\ell}} = 0$, SR invariance is restored, even when the Higgs field is still condensed $\braket{\vec{H}_{\vec{R},\ell}} \neq 0$, which defines a phase with topological order. Despite the preserved SR symmetry, discrete symmetries can still be broken \cite{chatterjeeIntertwiningTopologicalOrder2017,PhysRevB.98.235126}. As the symmetry analysis is mathematically identical to studying whether the associated classical magnetic orders preserve a symmetry modulo SRs, we conclude that the associated doped spin liquids have the same symmetries as those listed in \tableref{table:classicalphases}. That is, in agreement with our Schwinger-Boson study, the fractionalized phases associated with A$_{1,2,3}$ will respect all symmetries, those with B$_{1,2}$ will be nematic, and the one for phase C will exhibit altermagnetic loop currents \cite{chatterjeeIntertwiningTopologicalOrder2017,PhysRevB.98.235126}. Notably, both approaches also yield the same invariant gauge groups. 

The gauge theory further allows one to compute the electronic spectral function in those phases \cite{OurPNAS,PhysRevX.8.021048}, which is obtained as a convolution of the chargon and spinon Green's functions, see \figref{fig:ElectronicSpectra}(c). Focusing for concreteness on the fractionalized phase of A$_{1}$, we compare the resulting spectral function with the one in the magnetically ordered phase in \figref{fig:ElectronicSpectra}(d). As a result of the restored SU(2) SR symmetry, the spectral function $A_{\vec{k}}(\omega)$ in the fractionalized phase does not depend on spin and yet has peaks centered around the same frequencies as the state with long-range AM order. This signals fractionalization since, for a given spin species, each peak in the right panel is associated with ``half an electron''. When plotting $A_{\vec{k}}(\omega)$ as a function of $\vec{k}$ at the Fermi level, i.e., $\omega = 0$, in \figref{fig:ElectronicSpectra}(e) we clearly see the characteristic $\vec{k}$-dependent splitting of the Fermi surfaces. The band splitting persists at energies away from the Fermi level, see \figref{fig:ElectronicSpectra}(f), which further reveals that there are additional (incoherent) high-energy features associated with the presence of spinons.   A combination of quantum oscillations (probing the chargons) and (spin-resolved) photoemission \cite{2024arXiv240913504L} [measuring $A_{\vec{k}}(\omega)$] can identify signatures of such a fractionalized state experimentally.

%==============================================
\vspace{0.5em}
\section{Conclusion}In summary, our work shows that driving altermagnets into the frustrated regime with competing couplings in scenarios where quantum fluctuations play an important role gives rise to very rich physics, such as orbital altermagnets, various interesting spin liquids, including states with altermagnetic correlations in the charge sector, as well as the emergence of spin--symmetric band splitting. Given the increasingly large number of candidate materials for altermagnets \cite{gao_ai-accelerated_2023}, which typically exhibit multiple complex moments in their unit cell, it seems plausible that altermagnetic systems with sufficient degree of frustration \cite{2023arXiv230914812X, Biniskos2022Mar, Surgers2024Aug} can be identified.

\begin{acknowledgments}
The authors acknowledge funding by the European Union (ERC-2021-STG, Project 101040651---SuperCorr). Views and opinions expressed are however those of the authors only and do not necessarily reflect those of the European Union or the European Research Council Executive Agency. Neither the European Union nor the granting authority can be held responsible for them. J.A.S is thankful for discussions with Sayan Banerjee, Shubhayu Chatterjee, Lucas Pupim and Vitor Dantas. M.S.S.~acknowledges discussions and a previous collaboration with Sayan Banerjee on altermagnetism.
\end{acknowledgments}

%\bibliography{draft_1}


\newpage 
\onecolumngrid
\begin{appendix}

\section{Classical Phase Diagram }\label{app:classicalpd}
As discussed in the main text, to study the ordered phases for 
$\mathcal{H}$ defined in \equref{eq:heisenbergmodel}, we start from the classical limit $S\rightarrow \infty$. For this, we consider the classical ansatz \eqref{eq:ansatzeclassical} for the two sub-lattices in the unit cell given by 
\begin{equation}
\begin{array}{l}
\vec{S}_{A,\vec{\alpha}}=\left(\cos (\vec{Q} \cdot \vec{R}), \sin (\vec{Q} \cdot \vec{R}), \eta_A\right)^T/\sqrt{1+\eta_A^2} \\
\vec{S}_{B,\vec{\alpha}}=\left(\cos (\vec{Q}\cdot \vec{R}-\varphi), \sin (\vec{Q} \cdot \vec{R}-\varphi), \eta_B\right)^T/\sqrt{1+\eta_B^2} 
\label{eq:ansatze2}
\end{array},
\end{equation}
where $\vec{\alpha}=\{Q_{x}, Q_{y}, \varphi, \eta_{A}, \eta_{B}\}$ is a set of variational parameters yet to be determined.
We determine the classical phase diagram by minimizing the classical energy $\mathcal{H}\left(\vec{S}_{A,\vec{\alpha}}\cdot \vec{S}_{B,\vec{\alpha}}\right)$ expression with respect to the parameters $\vec{\alpha}$, using a combination of grid search over the variational parameters' space plus gradient descent-like optimization methods to find energetically favored configurations. 

The first step for the grid search method consists of defining an energy grid with a total of $L_{\text{GD}}$ equally spaced values for each of the variational parameters in $\vec{\alpha}$, with the domains for the initial values $\vec{\alpha}_{0}$ defined as 
$Q_{x}^{0}, Q_{y}^{0}, \varphi^{0} \in [-\pi, \pi]$ and  $\eta_{A}^{0}, \eta_{B}^{0} \in [-10, 10]$. 
For each pair of coupling constants $(J_{2}/J_{1}, KS^{2}/J_{1})$, we identify the configurations $\vec{\alpha}_{0}$ which have the lowest energy within the computed energy grid. These configurations are then selected as a starting point for the gradient descent algorithm. These set of parameters $\vec{\alpha}_{0}$ are iteratively updated with the standard formula 
$
\vec{\alpha} = \vec{\alpha}_{0} - \lambda \partial \mathcal{H}/\partial \vec{\alpha}.
$
Here, $\lambda$ is the learning rate, and each iteration constitutes one epoch. Our simulations included $\lambda \in [0.1, 0.001]$ over different runs up to $4\times10^{4}$ epochs. We repeated this process for varying energy grid sizes $L\in [10, 40]$.  Additionally, we also considered one variation of the standard gradient-descent method called ADAM, which introduces adaptable learning rates for each of the variational parameters \cite{kingmaAdamMethodStochastic2017}. After convergence, the minimum energy configurations at each $(KS^{2}/J_{1}, J_{2}/J_{1})$ point are characterized by the sets $\vec{\alpha}_{\text{conv}}$. Their energies were then compared to determine the phase boundaries in \figref{fig:classicalPD}(b). A particular set for each of the phases in the classical phase diagram (b) can be seen in \tableref{table:classicalphases}, as well as their spin textures $\vec{S}_{A,B}\left(\vec{\alpha}_{\text{conv}}\right)$ in \figref{fig:classicalPD}(c).

\begin{figure}[htp!]
\includegraphics[width=\linewidth]{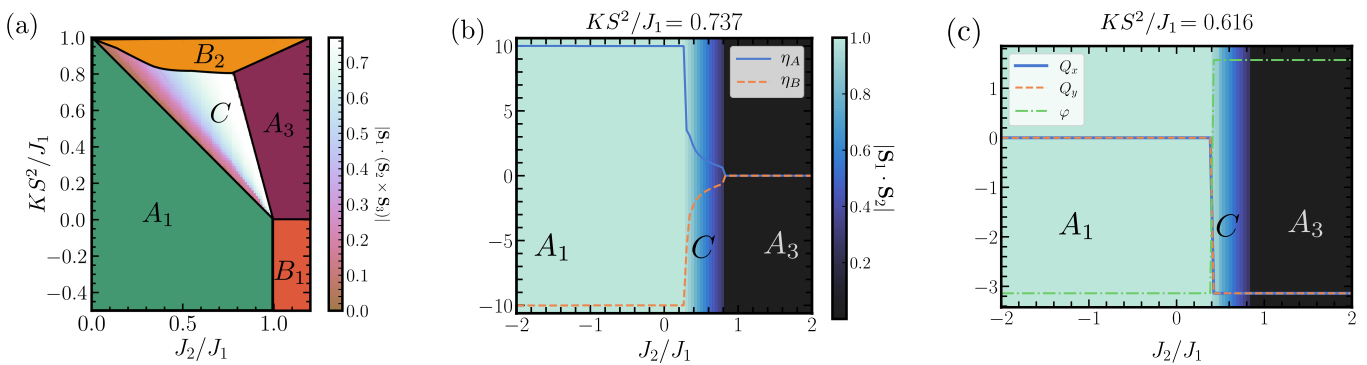}
    \caption{Phase diagram of the model \eqref{eq:heisenbergmodel} in the classical limit $S\rightarrow \infty$, focused around the region where the orbital AM ($C$) is located. (b) and (c) show one-dimensional cuts for the phase transition $A_{1}\rightarrow C \rightarrow A_{3}$ from the perspective of the variational parameters $\vec{\alpha}_{\text{conv}}$. The colormap in (a) refers to the absolute value of the scalar spin chirality $|\vec{S}_{1}\cdot\left(\vec{S}_{2}\times \vec{S}_{3}\right)|$,  whereas the background color in (b) and (c) to the absolute value of the nearest-neighbor spin product $|\vec{S}_{1}\cdot \vec{S}_{2}|$.}
    \label{fig:classicalPDzoom}
\end{figure}
As stated in the main text, each magnetic phase can also be characterized by two sets of observables: the nearest-neighbor spin products on the elementary plaquette (EP) given by $\vec{\mathcal{S}}^{\text{EP}} $ and the scalar spin chiralities  $\vec{\chi}^{\text{EP}}$. The latter are crucial for characterizing non-coplanar textures \cite{chatterjeeIntertwiningTopologicalOrder2017,messioLatticeSymmetriesRegular2011, hickeyEmergenceChiralSpin2017}. 
The orbital AM $C$ is the only phase in the classical phase diagram with non-zero scalar spin chirality. It has a staggered pattern $\vec{\chi}^{\text{EP}}$ (see \tableref{table:classicalphases}) that evolves continuously for varying $\left(KS^{2}/J_{1}, J_{2}/J_{1}\right)$, as can be seen in \figref{fig:classicalPD} (e) and \figref{fig:classicalPDzoom} (a). The continuous change of this observable is accompanied by a similar behavior from the perspective of the variational parameters $\eta_{A}$ and $\eta_{B}$ obtained in $\vec{\alpha}_{\text{conv}}$, which always obey the constraint $\eta_{A}+\eta_{B}=0$, see \figref{fig:classicalPDzoom}(b). Likewise, the nearest-neighbor spin products $\vec{\mathcal{S}}^{\text{EP}}$ also vary continuously in $C$ and $A_{2}$, see \figref{fig:classicalPD}(b, d) and background color in \figref{fig:classicalPDzoom}(b, c).      
The other phases $A_{1}$, $B_{1}$, $B_{2}$ and $A_{3}$ have fixed values for $\vec{\mathcal{S}}^{\text{EP}}$ and zero scalar spin chirality (see \tableref{table:classicalphases}).

\section{Schwinger-Boson ansätze}\label{app:SchwingerBoson}
\subsection{General formalism}
In this section we discuss in more detail the SB description of the spin-liquid states considered in the main text, as well as their associated magnetic orders obtained after the bosonic condensation. We start by rewriting the Hamiltonian \eqref {eq:heisenbergmodel}, without the ring exchange term $K$, in terms of the operators
\begin{equation}
\hat{\vec{S}}_{i}= \dfrac{1}{2} \sum_{\sigma, \sigma '=\uparrow, \downarrow} \hat{b}^{\dagger}_{i \sigma} {\vec{\sigma}}_{\sigma \sigma '} \hat{b}_{i \sigma '}. \label{RewritingOfTheSpin}
\end{equation}
The inclusion of only $J_{1}$ and $J_{2}$ exchange interactions in the SB formalism turns out to be sufficient to identify ansätze corresponding to all the magnetic phases present in the classical phase diagram in \figref{fig:classicalPD}(b). Using the SU$(2)$ completeness relation 
\begin{equation}
{\vec{\sigma}}_{\mu \mu '} \cdot {\vec{\sigma}}_{\nu \nu '} =\sum_{k=1}^3 \sigma_{\mu \mu '}^k \sigma_{\nu \nu '}^k =2{\delta}_{\mu \nu '} {\delta}_{\mu ' \nu} - {\delta}_{\mu \mu '} {\delta}_{\nu \nu '},  
\end{equation}
we can rewrite the dot product of spins as 
$\hat{\vec{S}}_i \cdot \hat{\vec{S}}_j = \mathcal{\hat{B}}^{\dagger}_{ij} \mathcal{B}_{ij} - \mathcal{\hat{A}}^{\dagger}_{ij} \mathcal{A}_{ij},
$
where $\mathcal{\hat{A}}_{ij}$ and $\mathcal{\hat{B}}_{ij}$ are bond operators defined as 
\begin{equation}
\mathcal{\hat{A}}_{ij}=\dfrac{1}{2} \sum_{\sigma, \sigma '=\uparrow, \downarrow} \hat{b}_{i \sigma} (i {\sigma}_y)_{ \sigma \sigma '} \hat{b}_{j \sigma} \quad \text{and} \quad \mathcal{\hat{B}}_{ij}=\dfrac{1}{2} \sum_{\sigma=\uparrow, \downarrow} \hat{b}_{i \sigma}^{\pdagger}  \hat{b}^{\dagger}_{j \sigma}. 
\label{eq:bonds}
\end{equation}
These are typically referred to as \textit{paring} and  \textit{hopping} operators, respectively.
After a Hubbard-Stratonovich transformation, we obtain the decoupled mean-field Hamiltonian
\begin{equation}\label{HMF1}
\mathcal{H}_{\text{MF}} =  \sum_{i,j,\sigma} \dfrac{1}{2} \big[ (\mathcal{B}^{\ast}_{ij} \hat{b}_{i \sigma} \hat{b}^{\dagger}_{j \sigma} - \mathcal{A}^{\ast}_{ij} \sigma \hat{b}_{i \sigma}  \hat{b}_{j - \sigma}  + \text{H.c.} ) + ( | \mathcal{A}_{ij}|^2 - | \mathcal{B}_{ij}|^2 )/J_{ij} \big] - \mu \sum_{i} \hat{b}^{\dagger}_{i \sigma} \hat{b}_{i \sigma}, 
\end{equation}
where $\mu$ is the chemical potential and $\mathcal{A}_{ij}$, $\mathcal{B}_{ij}$ are the mean-field values of the bond operators, i.e., 
\begin{equation}
\langle \mathcal{\hat{A}}_{ij} \rangle = \mathcal{A}_{ij}/J_{ij} \quad \text{and} \quad \langle \mathcal{\hat{B}}_{ij} \rangle = \mathcal{B}_{ij}/J_{ij}.   
\end{equation}
Equation (\ref{HMF1}) has the same functional form as the Hamiltonian $\mathcal{H}_{b}$ mentioned in the main text, apart from the  constant energy shift $\mathcal{H}_{\text{MF}}^{0}=\sum_{i,j\sigma}( | \mathcal{A}_{ij}|^2 - | \mathcal{B}_{ij}|^2 )/J_{ij} $. 
The mean-field parameters $\mathcal{A}_{ij}$, $\mathcal{B}_{ij}$ can assume complex values and satisfy the constraints
$
\mathcal{A}_{ij}=-\mathcal{A}_{ji},$ and $ \mathcal{B}_{ij}=\mathcal{B}^{\ast}_{ji}$.
The Hamiltonian (\ref{HMF1}) exhibits an emergent U$(1)$ gauge symmetry, which can be clearly seen by  noticing that $\mathcal{H}_{\text{MF}}$ remains invariant under a local U(1) gauge transformation defined by
\begin{equation}
\begin{split}
\hat{b}_{i \sigma} & \rightarrow e^{i {\phi}_i} \hat{b}_{i \sigma}, \quad \sigma=\uparrow, \downarrow \\
\mathcal{A}_{ij} & \rightarrow e^{i ({\phi}_i + {\phi}_j)} \mathcal{A}_{ij}, \\
\mathcal{B}_{ij} & \rightarrow e^{i ({\phi}_i - {\phi}_j)} \mathcal{B}_{ij}.
\end{split}
\end{equation}

Consequently, each spin-liquid ansatz can be identified through gauge-invariant fluxes defined as 
\begin{equation}
\begin{split}
{\Phi}^{{\Box}_{\mathcal{AA}}}_{ijkl} &= \text{Arg}[\mathcal{A}_{ij}\mathcal{A}^{\ast}_{jk}\mathcal{A}_{kl}\mathcal{A}^{\ast}_{li}]; \\
{\Phi}^{{\Box}_{\mathcal{AB}}}_{ijkl} &= \text{Arg}[\mathcal{A}_{ij}\mathcal{A}^{\ast}_{jk}\mathcal{B}_{kl}\mathcal{B}_{li}]; \\
\widetilde{{\Phi}}^{{\Box}_{\mathcal{AB}}}_{ijkl} &= \text{Arg}[\mathcal{A}_{ij}\mathcal{B}^{\ast}_{jk}\mathcal{A}^{\ast}_{kl}\mathcal{B}_{li}]; \\
{\Phi}^{\Delta}_{ijk} &= \text{Arg}[\mathcal{A}_{ij}\mathcal{A}^{\ast}_{jk}\mathcal{B}_{ki}]; \\
{\Phi}^{{\Delta}_2}_{ijk} &= \text{Arg}[\mathcal{B}_{ij}\mathcal{A}_{jk}\mathcal{A}^{\ast}_{ki}] .
\end{split}
\end{equation}
We will henceforth refer to each ansatz according to the notation $[ {\Phi}^{{\Box}_{\mathcal{AA}}}_{1234}, {\Phi}^{{\Box}_{\mathcal{AB}}}_{1234}, \widetilde{{\Phi}}^{{\Box}_{\mathcal{AB}}}_{1234}, \{ {\Phi}^{\Delta}_{123} , {\Phi}^{\Delta}_{234}, {\Phi}^{\Delta}_{341}, {\Phi}^{\Delta}_{412} \},$ $ \{ {\Phi}^{{\Delta}_2}_{123}, {\Phi}^{{\Delta}_2}_{234}, {\Phi}^{{\Delta}_2}_{341} , {\Phi}^{{\Delta}_2}_{412} \}]$.

Focusing on ansätze and gauges where the $\mathcal{A}_{ij}$, $\mathcal{B}_{ij}$ respect the Bravais-lattice translational symmetry, the Hamiltonian~\eqref{HMF1} can be rewritten in momentum space after taking the Fourier transformation 
$
\hat{b}_{\vec{r},  \alpha, \sigma } = \dfrac{1}{\sqrt{N}} \sum_{{\vec{k}} } e^{-i \vec{k} . \vec{r}} \hat{b}_{\vec{k}, \alpha, \sigma},
$ as
$
\mathcal{H}_{\text{MF}} = \sum_{\vec{k}}  {\Psi}^{\dagger}_{\vec{k}} M_{\vec{k}}\left(\mathcal{A}_{ij}, \mathcal{B}_{ij}\right) {\Psi}_{\vec{k}} 
+\textit{const}$. Here, ${\Psi}_{\vec{k}} = \left(\hat{b}^{\pdagger}_ {\vec{k} A \uparrow}, \hat{b}^{\pdagger}_ {\vec{k} B \uparrow}, \hat{b}^{\dagger}_ {-\vec{k} A \downarrow}, \hat{b}^{\dagger}_ {-\vec{k}B \downarrow} \right)^T$ is the Nambu spinor and the $4\times 4$ matrix $M_{\vec{k}}$ is given by  
\begin{equation}
\footnotesize
M_{\vec{k}}\left(\mathcal{A}_{ij}, \mathcal{B}_{ij}\right)=\dfrac{1}{2}
\begin{pmatrix}
\mathcal{B}^{\ast}_{24} e^{- i k_x} + \mathcal{B}_{24} e^{i k_x} - 2\mu  &  {\tilde{\xi}}^{\ast}_{\vec{k}} & -2 i \mathcal{A}_{24} \sin(k_x) & -{\xi}_{-{\vec{k}}} \\
{\tilde{{\xi}}}_{\vec{k}} & \mathcal{B}^{\ast}_{13} e^{i k_y} + \mathcal{B}_{13} e^{- i k_y} - 2\mu  &  {\xi}_{\vec{k}}  &  2 i \mathcal{A}_{13} \sin(k_y) \\
 2 i \mathcal{A}^{\ast}_{24} \sin(k_x) &  {\xi}^{\ast}_{\vec{k}} & \mathcal{B}^{\ast}_{24} e^{ i k_x} + \mathcal{B}_{24} e^{-i k_x} - 2\mu  & {\tilde{{\xi}}}_{-{\vec{k}}} \\
-{\xi}^{\ast}_{-{\vec{k}}} & - 2 i \mathcal{A}^{\ast}_{13} \sin(k_y)  &  {\tilde{{\xi}}}^{\ast}_{-{\vec{k}}} &  \mathcal{B}^{\ast}_{13} e^{- i k_y} + \mathcal{B}_{13} e^{i k_y} - 2\mu
\end{pmatrix}\label{MMatrixRep}
\end{equation}
with  ${\tilde{{\xi}}}_{\vec{k}} = \mathcal{B}^{\ast}_{23} + \mathcal{B}_{12} e^{-i k_y} + \mathcal{B}_{34} e^{i k_x} + \mathcal{B}_{14} e^{-i(k_y-k_x)} $ and ${\xi}_{\vec{k}}= -\mathcal{A}_{12}e^{- i k_y} + \mathcal{A}_{23} - \mathcal{A}_{34}e^{i k_x} - \mathcal{A}_{14}e^{- i (k_y-k_x)}$.

\subsubsection{Diagonalization of the quadratic bosonic Hamiltonians}
In order to diagonalize $\mathcal{H}_{\text{MF}}$, we consider the Bogoliubov-Valatin transformation, a well-known method for diagonalizing a general quadratic bosonic Hamiltonian \cite{Bogolyubov:1947zz,valatin1958comments,COLPA1986417,xiao2009}. For simplicity, we suppress the momentum index $\vec{k}$ for now and consider a Hamiltonian $H$ of the form
\begin{equation}
H=\dfrac{1}{2} {\Psi}^{\dagger} M_{\vec{k}} \Psi \quad \text{with}\quad \ \ {\Psi}^{\dagger}=\left(\hat{b}^{\dagger}_1, \ldots , \hat{b}^{\dagger}_N, \hat{b}_1, \ldots , \hat{b}_N\right)^{T},
\end{equation}
with $N$ different values of internal quantum numbers describing degrees of freedom such as spin or sublattice, for example. We then reformulate the bosonic Hamiltonian $H$ in terms of a new set of creation and annihilation operators (${\hat{\gamma}}^{\dagger}_m$ and ${\hat{\gamma}}_m$). The diagonalizability of the original bosonic Hamiltonian $H$ ensures the existence of such a transformation matrix $T$ that turns $M_{\vec{k}}$ into a simplified diagonal-matrix form in 
\begin{equation}
H = \dfrac{1}{2} \ {\Gamma}^{\dagger} T^{\dagger} M_{\vec{k}} T {\Gamma} \quad \text{with} \quad
T^{\dagger}M_{\vec{k}}T =\begin{pmatrix}
{\omega}_1 & 0 & \dots & 0 \\
0 & {\omega}_2 & \dots & 0 \\
\vdots & \vdots & \ddots & \vdots\\
0 & 0 & \dots & {\omega}_{2N} 
\end{pmatrix}, \quad 
\Psi = T \Gamma \quad \text{and} \ \ \ {\hat{\Gamma}^{\dagger} \equiv \left({\hat{\gamma}}^{\dagger}_1, \ldots , {\hat{\gamma}}^{\dagger}_N, {\hat{\gamma}}_1, \ldots , {\hat{\gamma}}_N\right)^{T}}.
\end{equation} 
The transformation matrix $T$  can be constructed from the eigenvectors ($V({\omega}_i)$)  of $ K_{\vec{k}} = {\rho}_3 M_{\vec{k}} $, known as the dynamic matrix. The eigenvalues of the dynamic matrix $K_{\vec{k}}$ appear in pairs and are real if $K_{\vec{k}}$ is diagonalizable. Furthermore, the pair of eigenvalues is generally related to two eigenvectors, one with a positive norm and the other with a negative norm, i.e.,  
\begin{equation}
V^{\dagger} ({\omega}_i) {\rho}_3 V({\omega}_i)=1 \quad \text{and} \quad V^{\dagger} (-{\omega}_i) {\rho}_3 V(-{\omega}_i)=-1, \quad  \text{where} \quad
{\rho}_3 = \begin{pmatrix}
{\mathbbm{1}}_{N \times N} & 0 \\
0 & - {\mathbbm{1}}_{N \times N}
\end{pmatrix}.
\end{equation}
Finally, bosonic commutation relations constrain the transformation matrix $T$ to be paraunitary, i.e.,
$
T {\rho}_3 T^{\dagger}={\rho}_3,
\label{eq:conditionparaunitary}
$
suggesting a natural order for grouping the eigenvectors in $T$  as
\begin{equation}
\label{eq:bogovala}
T=\big[ V({\omega}_1), \ldots , V({\omega}_N), V(-{\omega}_1), \ldots , V(-{\omega}_N)  \big],
\quad \text{yielding} \quad
\begin{cases}
T^{-1}K_{\vec{k}}T = \text{diag} ({\omega}_1, \ldots, {\omega}_N, -{\omega}_1, \ldots, -{\omega}_N),\\ 
T^{-1}M_{\vec{k}}T = \text{diag} ({\omega}_1, \ldots, {\omega}_N, {\omega}_1, \ldots, {\omega}_N) .
\end{cases}
\end{equation}  

\subsubsection{Determining the magnetic order}
In the SB formalism, bosonic condensation occurs at specific momentum points $\pm \vec{k}_0$, corresponding to the minima of the spinon bands, which are obtained from the diagonalization of $K_{\vec{k}}$. At a critical chemical potential $\mu_{c}$, these points are associated with zero-energy eigenvectors $\Psi_{i}$. In real space, the condensate can then be expressed as a linear combination of these eigenvectors as (assuming for concreteness that $\vec{k}_0 \neq -\vec{k}_0$)
\begin{equation}
\langle \Psi \rangle = \left(\langle \hat{b}_{\vec{r}A \uparrow}^{\pdagger} \rangle  , \langle \hat{b}_{\vec{r}B \uparrow}^{\pdagger}  \rangle, \langle \hat{b}_{\vec{r}A \downarrow}^{\dagger} \rangle , \langle \hat{b}_{\vec{r}B \downarrow}^{\dagger}  \rangle  \right)^{T}= e^{i \vec{k}_0 \cdot \vec{r}} (z_1  {\Psi}_1 + z_2  {\Psi}_2) + e^{-i \vec{k}_0 \cdot \vec{r}} (z_3  {\Psi}_3 + z_4  {\Psi}_4) \quad \text{with} \quad  z_i \in \mathbb{C}.
\label{eq:SU(2)eqn} 
\end{equation}
The expectation value of the spins at positions $\vec{r}$ (spin texture) can then be obtained from 
\begin{equation}
\langle \vec{S}_{\mu}(\vec{r}) \rangle = X^{\dagger}_{\mu } \vec{\sigma }X_{\mu},
\label{eq:cond2}
\end{equation}
up to global SR transformations, where  $X_{\mu}=\left( \langle \hat{b}_{\vec{r}\mu \uparrow} \rangle, \langle \hat{b}_{\vec{r}\mu \downarrow} \rangle\right)^T$. Here, $\mu$ refers to the sublattice indices, and $\vec{\sigma} = \left(\sigma_x,\sigma_y,\sigma_z\right)^T$. Finally, spin liquid ansätze can be directly associated to classically magnetic ordered phases using these expressions. In what follows, we outline the key intermediate steps in employing this procedure for the ansätze in \figref{fig:SchwingerBoson} and the magnetic phases in \figref{fig:classicalPD}(b, c):

\begin{itemize}
\item $A_1$ phase - U$(1)$ $[0,0,0, \{0,0,0, 0\}, \{0,0,0,0\}]$:

The gauge-flux structure is given by only non-zero nearest-neighbor bonds $\mathcal{A}_{ij}$ [see \figref{fig:SchwingerBoson}]:
\begin{equation}
\begin{split}
& \mathcal{A}_{12}=\mathcal{A}_1, \
\mathcal{A}_{23}=\mathcal{A}_1, \
\mathcal{A}_{34}=-\mathcal{A}_1,  \
\mathcal{A}_{41}=-\mathcal{A}_1; \\& 
\mathcal{A}_{13}=0, \ \mathcal{A}_{24}=0; \\
& \mathcal{B}_{12}=0, \ \mathcal{B}_{23}=0, \ \mathcal{B}_{34}=0, \ \mathcal{B}_{41}=0; \\
&\mathcal{B}_{13}=0, \ \mathcal{B}_{24}=0.
\end{split}
\end{equation}
For this state, we find four-fold degenerate spinon dispersions (related to $M_{\vec{k}}$, see \eqref{eq:bogovala})
\begin{equation}
{\omega} =  \sqrt{{\mathcal{A}_1}^2 f_{1}(\vec{k})+\mu ^2} \quad \text{with} \quad  f_{1}(\vec{k})=2\cos \left(k_{x}\right)\sin^2 \left(k_{y}/2\right)+\cos \left(k_{y}\right)
\end{equation}
The spinon condensation occurs at $\vec{Q}=(0,\pi)$ as the chemical potential approaches ${\mu}_c=  2  \mathcal{A}_1$. There are four zero-energy eigenvectors, which are given by
\begin{equation}
\begin{split}
  & \psi_{1} =  \frac{1}{N_{\psi_{1}}}\left(\frac{\mu_{c}-\sqrt{-4\mathcal{A}_{1}^{2}+\mu_{c}^{2}}}{2\mathcal{A}_{1}}, 0, 0, 1\right)^{T} \;, \; \psi_{2}= \frac{1}{N_{\psi_{2}}}\left(\frac{\mu_{c}+\sqrt{-4\mathcal{A}_{1}^{2}+\mu_{c}^{2}}}{2\mathcal{A}_{1}}, 0, 0, 1\right)^{T},\; \\ 
  & \psi_{3} =  \frac{1}{N_{\psi_{3}}}\left(0, -\frac{\mu_{c}+\sqrt{-4\mathcal{A}_{1}^{2}+\mu_{c}^{2}}}{2\mathcal{A}_{1}}, 1, 0\right)^{T} \; \text{and} \; \psi_{4}= \frac{1}{ N_{\psi_{4}}}\left(0, \frac{-\mu_{c}-\sqrt{-4\mathcal{A}_{1}^{2}+\mu_{c}^{2}}}{2\mathcal{A}_{1}}, 1, 0\right)^{T},\; 
\end{split}
\end{equation}
where $N_{\psi_{i}}$ represents the  corresponding normalization factors for each eigenvector.  For the cases where we do not analytically diagonalize $K_{\vec{k}}$, condensation is reached numerically by introducing a small offset in $\mu\rightarrow \mu_{c} +\Delta \mu$, where $\Delta \mu\approx 10^{-10}$.  After considering equations \eqref{eq:SU(2)eqn} and \eqref{eq:cond2}, we calculate the observables $\vec{\mathcal{S}}^{\text{EP}}$ and 
$\vec{\mathcal{\chi}}^{\text{EP}}$ and conclude that 
the resulting magnetic order corresponds to the AM $A_{1}$ in \figref{fig:classicalPD}(e).

\item $A_2$ phase -  $Z_{2}$ $[0,\pi ,0, \{0,0,0, 0\}, \{0,0,0,0\}]$:  

This ansatz is obtained from the previous U$(1)$ case by turning on the nearest-neighbor bonds $\mathcal{B}_{ij}$ as [see \figref{fig:SchwingerBoson}]:
\begin{equation}
\begin{split}
%& \mathcal{A}_{12}=\mathcal{A}_1, 
%\mathcal{A}_{23}=-\mathcal{A}_1, 
%\mathcal{A}_{34}=-\mathcal{A}_1, 
% \mathcal{A}_{41}=\mathcal{A}_1; \\
% & \mathcal{A}_{13}=0, \ 
% \mathcal{A}_{24}=0; \\
% & \mathcal{B}_{12}=i \mathcal{B}_1, \
% \mathcal{B}_{23}=-i \mathcal{B}_1, \
% \mathcal{B}_{34}=-i \mathcal{B}_1, \
% \mathcal{B}_{4}=i \mathcal{B}_1; \ \\
% &\mathcal{B}_{13}=0, \
% \mathcal{B}_{24}=0;
& \mathcal{A}_{12}=\mathcal{A}_1, 
\mathcal{A}_{23}=\mathcal{A}_1, 
\mathcal{A}_{34}=-\mathcal{A}_1, 
\mathcal{A}_{41}=-\mathcal{A}_1; \\
& \mathcal{A}_{13}=0, \ 
\mathcal{A}_{24}=0; \\
& \mathcal{B}_{12}=i \mathcal{B}_1, \
\mathcal{B}_{23}=i \mathcal{B}_1, \
\mathcal{B}_{34}=-i \mathcal{B}_1, \
\mathcal{B}_{41}=-i \mathcal{B}_1; \ \\
&\mathcal{B}_{13}=0, \
\mathcal{B}_{24}=0.
\end{split}
\end{equation}
The corresponding dispersions are given by  
\begin{equation}
\omega_{\pm} = \sqrt{\left(\mathcal{A}_{1}^{2}-\mathcal{B}_{1}^{2}\right) f_{2}(\vec{k})+ \mu^{2}\pm |\mathcal{B}_{1}g(\vec{k})|\sqrt{A^{2}f_{2}(\vec{k})+\mu^{2}}},
\end{equation}
with $f_{2}(\vec{k}) = 2 \cos(k_{y}) \sin^{2}\left(k_{x}/2\right)+\cos \left(k_{x}\right)$ and $g\left(\vec{k}\right)=4\sin \left(k_{x}/2\right)\cos \left(k_{y}/2\right)$.  The spinon condensation takes place at the point $\vec{Q}=\left(\pi, 0\right)$ as the chemical potential approaches $\mu_{c} = - 2\sqrt{\mathcal{A}_{1}^{2}+\mathcal{B}_{1}^{2}}$. The associated zero-energy eigenvectors to the condensation are expressed as
\begin{equation}
\begin{split}
  & \psi_{1} =  \frac{1}{N_{\psi}}\left(-\frac{\sqrt{\mathcal{A}_{1}^{2}+\mathcal{B}_{1}^{2}}}{\mathcal{A}_{1}},  i \frac{\mathcal{B}_{1}}{\mathcal{A}_{1}},0,1\right)^{T} \; \text{and} \; \psi_{2}=\frac{1}{N_{\psi}}\left( i \frac{\mathcal{B}_{1}}{\mathcal{A}_{1}},\frac{\sqrt{\mathcal{A}_{1}^{2}+\mathcal{B}_{1}^{2}}}{\mathcal{A}_{1}},1,0\right)^{T}, \;
\end{split}
\end{equation}
with $N_{\psi}$ representing the normalization factor.
The resulting magnetic order associated to the condensation is the canted AM $A_{2}$ phase in \figref{fig:classicalPD}(e).

\item $B_1$ phase - U$(1)$ $[0,0,0, \{0,0,0, 0\}, \{0,0,0,0\}]$:

The gauge-flux structure is given by real nearest-neighbor and next-nearest-neighbor bonds $\mathcal{A}_{ij}$ [see \figref{fig:SchwingerBoson}]:
\begin{equation}
\begin{split}
& \mathcal{A}_{12}=\mathcal{A}_1, \ 
\mathcal{A}_{23}=0,\ 
\mathcal{A}_{34}=-\mathcal{A}_1, \
\mathcal{A}_{41}=0;\\ 
&\mathcal{A}_{13}=\mathcal{A}_2, \ \mathcal{A}_{24}=-\mathcal{A}_2; \\
& \mathcal{B}_{12}=0, \ 
\mathcal{B}_{23}=0, \ 
\mathcal{B}_{34}=0, \ 
\mathcal{B}_{41}=0; \\
& \mathcal{B}_{13}=0, \ \mathcal{B}_{24}=0.
\end{split}
\end{equation}
This leads to doubly degenerate bands whose dispersions are:
\begin{equation}
\begin{split}
& {\omega}_{\pm}  = \frac{1}{2}\sqrt{2\mathcal{A}_1^2f_{3}\left(\vec{k}\right)+4\mu^{2}-2\mathcal{A}_{2}^{2}f_{4}\left(\vec{k}\right)\pm \left| \mathcal{A}_2 m_{+}\left(\vec{k}\right)\right| \sqrt{-8 \mathcal{A}_1^2f_{3}\left(\vec{k}\right)-2\mathcal{A}_{2}^{2}m_{-}^{2}\left(\vec{k}\right)}},
\end{split}
\end{equation}
with $f_{3}\left(\vec{k}\right)=\cos\left(k_{x}+k_{y}\right)-1$, $f_{4}\left(\vec{k}\right)=\sin ^{2}\left(k_{x}\right)+\sin^{2} \left(k_{y}\right)$ and  $m_{\pm}\left(\vec{k}\right)=\sin\left(k_{x}\right)\pm\sin\left(k_{y}\right)$. 
Bosonic condensation takes place at ${\mu}_c= \mathcal{A}_1 + \mathcal{A}_2$ and $\vec{Q}=\pm \left(\pi/2, \pi/2\right)$. Approaching the condensation numerically with $\mathcal{A}_1=-0.2$ and $\mathcal{A}_2=-0.5$, the zero-energy eigenvectors at $\vec{Q}$ are given by 
$ \psi_{1} =  \frac{1}{2}\left(1,1,i, i\right)^{T} \; \text{and} \; \psi_{2} =\frac{1}{2}\left(-i,-i,1, 1\right)^{T} \;
$, with $B_{1}$ being the associated magnetic order.

\item $B_2$ phase - U$(1)$ $[0,0,0, \{0,0,0, 0\}, \{0,\pi,0,0\}]$:

The gauge-flux structure is given by both real nearest-neighbor and next-nearest-neighbor bonds $\mathcal{A}_{ij}$ and $\mathcal{B}_{ij}$ [see \figref{fig:SchwingerBoson}]:
\begin{equation}
\begin{split}
& \mathcal{A}_{12}=0, \
\mathcal{A}_{23}=0, \
\mathcal{A}_{34}=-\mathcal{A}_1, \
\mathcal{A}_{41}=-\mathcal{A}_1; \\
& \mathcal{A}_{13}=0, \
\mathcal{A}_{24}=\mathcal{A}_2; \\
& \mathcal{B}_{12}= \mathcal{B}_1, \
\mathcal{B}_{23}=\mathcal{B}_1, \
\mathcal{B}_{34}=0, \
\mathcal{B}_{41}=0; \\
& \mathcal{B}_{13}=\mathcal{B}_2, \
 \mathcal{B}_{24}=0.
\end{split}
\end{equation}
It leads to doubly degenerate bands. The spinon condensation takes place at $\vec{Q}= \pm \left( \pi/2, 0 \right)$ as we approach the critical chemical potential given by ${\mu}_c = \frac{1}{2} (\mathcal{A}_2 + \mathcal{B}_2) - \frac{1}{2}\sqrt{\mathcal{A}_2^2 + 4 \mathcal{B}_1^2 - 2\mathcal{A}_2 \mathcal{B}_2 +\mathcal{B}_2^2 } $. Approaching the condensation numerically with $\mathcal{A}_1=-0.12$, $A_2=-0.35$, $\mathcal{B}_1=-0.6 $ and $\mathcal{B}_2=0.3$, we find the eigenvectors at $\vec{Q}$ to be given by $\psi_{1} = \left(0.60752, 0.36185, 0.60750 i, 0. + 0.36183 i\right)^{T} $ and $ \psi_{2}= \left({- 0.6075 i, - 0.36183 i, 0.60752, 0.36185}\right)^{T}$. The resulting magnetic order associated to the condensation is the $B_2$ phase in \figref{fig:classicalPD}(c).  

\item $A_3$ phase - $Z_2[\pi,0,0, \{ 0,0,0,0 \} , \{ 0,0,0,0\} ]$: 

The gauge-flux structure is given by real nearest-neighbor and next-nearest-neighbor bonds $\mathcal{A}_{ij}$ [see \figref{fig:SchwingerBoson}]:
\begin{equation}
\begin{split}
& \mathcal{A}_{12}=-\mathcal{A}_1, \
\mathcal{A}_{23}=\mathcal{A}_1,\
 \mathcal{A}_{34}=-\mathcal{A}_1,\
\mathcal{A}_{41}=-\mathcal{A}_1,\  \\
& \mathcal{A}_{13}=-\mathcal{A}_2, \mathcal{A}_{24}=-\mathcal{A}_2; \\
& \mathcal{B}_{12}=0, \mathcal{B}_{23}=0, \mathcal{B}_{34}=0, \ \mathcal{B}_{14}=0; \\
& \mathcal{B}_{13}=0, \mathcal{B}_{24}=0.
\end{split}
\end{equation}
In this case, we discover the spinon bands to be doubly degenerate bands, given by
\begin{equation}
\begin{split}
& {\omega}_{\pm}  = \frac{1}{2}\sqrt{4\mathcal{A}_1^2f_{5}\left(\vec{k}\right)+4\mu^{2}-2\mathcal{A}_{2}^{2}f_{4}\left(\vec{k}\right)\pm \sqrt{2} \left| \mathcal{A}_2 m_{-}\left(\vec{k}\right)\right| \sqrt{-8 \mathcal{A}_1^2f_{5}\left(\vec{k}\right)+ 2\mathcal{A}_{2}^{2}m_{+}^{2}\left(\vec{k}\right)}},
\end{split}
\end{equation}
with $f_{5}\left(\vec{k}\right)=\sin \left(k_{x}\right)\sin \left(k_{y}\right)-1$, $f_{4}\left(\vec{k}\right)=\sin ^{2}\left(k_{x}\right)+\sin^{2} \left(k_{y}\right)$ and  $m_{\pm}\left(\vec{k}\right)=\sin\left(k_{x}\right)\pm\sin\left(k_{y}\right)$. The minimum of the dispersion appears at  $\pm Q=\pm \left(-\pi/2,\pi/2 \right)$, with critical chemical potential given by ${\mu}_c=-\left|\sqrt{2} \mathcal{A}_1 + \mathcal{A}_2\right|$. For $\mathcal{A}_{1}=-0.3$ and $\mathcal{A}_{2}=-0.5$, the normalized eigenvectors are given by
 $\psi_{1}= \left(0.353553\left(1-i\right),-0.5i,0.353554 \left(1+i\right), 0.5 \right)^{T} $ and $ \psi_{2}= \left(0.5, 0.353554 \left(1-i\right), 0.5i, 0.353553 \left(1+i\right)\right)^{T} $. The resulting magnetic order associated to the condensation is the  $A_{3}$ phase in \figref{fig:classicalPD}(c).

\item $C$ phase - $Z_2[\pi, \pi, \pi, \{ \dfrac{3\pi}{2}, \dfrac{\pi}{2}, \dfrac{3\pi}{2} ,\dfrac{\pi}{2} \} , \{ \dfrac{3\pi}{2}, \dfrac{\pi}{2}, \dfrac{3\pi}{2} ,\dfrac{\pi}{2} \} ]$.  

Finally, the ansatz related to the orbital AM $C$ takes into account all types of bonds $\mathcal{A}_{ij}$ and $\mathcal{B}_{ij}$, i.e., 
\begin{equation}
\begin{split}
& \mathcal{A}_{12}=-\mathcal{A}_1, \ \mathcal{A}_{23}=\mathcal{A}_1, \ \mathcal{A}_{34}=-\mathcal{A}_1, \ \mathcal{A}_{41}=-\mathcal{A}_1; \\ 
& \mathcal{A}_{13}=-\mathcal{A}_2, \ \mathcal{A}_{24}=-\mathcal{A}_2; \\
& \mathcal{B}_{12}= i \mathcal{B}_1,  \ \mathcal{B}_{23}= i \mathcal{B}_1,  \ \mathcal{B}_{34}= i \mathcal{B}_1, \mathcal{B}_{41}= -i \mathcal{B}_1; \\ & 
\mathcal{B}_{13}= i \mathcal{B}_2, \mathcal{B}_{24}=-i \mathcal{B}_2.
\end{split}
\end{equation}
This results in two doubly degenerate bands which display minima at $\vec{Q}=\pm \left(-\pi/2, \pi/2 \right)$. For $\mathcal{A}_1=-0.3, \mathcal{A}_2=-0.2, \mathcal{B}_1=-0.35$ and $\mathcal{B}_2=-0.1$, the zero-energy eigenvectors are given by $ \psi_{1}=\left(0.41574 + 0.17646 i, 0.54409, 
 0.38472 + 0.38472 i, -0.41874 + 0.16920 i\right)^{T} $ and $\psi_{2} =\left({0.41874 - 0.16919 i, 
 0.38472 - 0.38472 i, 0.54409, -0.17646 + 0.41574 i}\right)^{T} $.

\end{itemize}
In addition, the spin correlations $\vec{\mathcal{S}}^{\text{EP}}$ and the spin scalar chiralities $\vec{\mathcal{\chi}}^{\text{EP}}$ can also be calculated in the spin liquid regime to be further compared with the same observables for the classically magnetic orders.  In the QSL regime, the expectation values are determined by evaluating all possible Wick contractions of the Schwinger boson operators in $ {\hat{\vec{S}}}_i \cdot {\hat{\vec{S}}}_j $ and $ {\hat{\vec{S}}}_i \cdot ( {\hat{\vec{S}}}_j \times {\hat{\vec{S}}}_k )$. Using equations \eqref{RewritingOfTheSpin} and \eqref{eq:bonds}, these are given by
\begin{equation}
\langle {\hat{\vec{S}}}_i \cdot {\hat{\vec{S}}}_j \rangle  = \frac{3}{2} \Big( \dfrac{|\mathcal{B}_{ij}|^2 - |\mathcal{A}_{ij}|^2}{J^2_{ij}} \Big),
\label{eq:ob1}
\end{equation} 
and 
\begin{equation}
\label{eq:ob2}
\langle {\hat{\vec{S}}}_i \cdot ( {\hat{\vec{S}}}_j \times {\hat{\vec{S}}}_k )  \rangle  = - \dfrac{3}{J_{ij}J_{jk}J_{ki}} \Big[  \text{Im} (\mathcal{B}_{ij}\mathcal{B}_{jk}\mathcal{B}_{ki}) + \text{Im} (\mathcal{A}_{ij}\mathcal{A}_{jk}^{\ast} \mathcal{B}_{ki}) + \text{Im} (\mathcal{B}_{ij}\mathcal{A}_{jk} \mathcal{A}_{ki}^{\ast} ) + \text{Im} (\mathcal{A}_{ij}^{\ast}\mathcal{B}_{jk} \mathcal{A}_{ki} ) \Big].
\end{equation}
These quantities in the quantum spin liquid (QSL) regime are gauge invariant, as expected. Moreover, equation (\ref{eq:ob2}) confirms that a mean-field ansatz based on purely real $\mathcal{A}_{ij}$ and $\mathcal{B}_{ij}$ is incapable of generating spin liquids with non-zero $\vec{\chi}^{\text{EP}}$, and, by extension, non-coplanar magnetic phases upon bosonic condensation.  More specifically, by using the ansatz  $Z_2[\pi, \pi, \pi, \{ \dfrac{3\pi}{2}, \dfrac{\pi}{2}, \dfrac{3\pi}{2} ,\dfrac{\pi}{2} \} , \{ \dfrac{3\pi}{2}, \dfrac{\pi}{2}, \dfrac{3\pi}{2} ,\dfrac{\pi}{2} \} ]$ we confirm that the expected staggered chirality pattern for phase $C$ can be found in both regimes, see Table \ref{table:SBMFTobs1}.
\begin{table*}[h!]
\begin{ruledtabular}
\caption{Values of the observables $\vec{\mathcal{S}}^{\text{EP}}$ and $\vec{\chi}^{\text{EP}}$ according to equations \eqref{eq:ob1} and \eqref{eq:ob2} before (in the quantum spin liquid regime) and after bosonic condensation (for the equivalent classical magnetic order).  The latter is calculated using equations \eqref{eq:SU(2)eqn} and \eqref{eq:cond2}. For the QSLs, the observables are presented up to a multiplicative factor. The non-universal parameters $s_{0}, s', s_{1},\chi_{1} \in \mathbb{R}^{+}$ depend on the mean-field values $\mathcal{A}_{ij}$ and $\mathcal{B}_{ij}$.}
\label{table:SBMFTobs1}
\centering
\begin{tabular}{c c c c c c} 
& \multicolumn{2}{c}{QSL} & \multicolumn{2}{c}{Condensate} \\ \cline{2-3}  \cline{4-5} 
 Label  & $\vec{\mathcal{S}}^{\text{EP}}$ & $\vec{{\chi}}^{\text{EP}} $ & $\vec{\mathcal{S}}^{\text{EP}}$  & $\vec{{\chi}}^{\text{EP}} $ \\ [0.5ex] 
 \hline
$A_1$ & $(-1,-1,-1,-1 )$  & $(0,0,0,0)$  & $(-1,-1,-1,-1 )$  & $(0,0,0,0)$  \\
$A_2$ & $ (-1,-1,-1,-1 )$  &  $(0,0,0,0)$  &  $s_0(-1,-1,-1,-1)$  & $(0,0,0,0)$ \\
$A_3$ &  $  (-1,-1,-1,-1 )$ & $(0,0,0,0)$  & $(0,0,0,0)$  &  $(0,0,0,0)$ \\
$B_1$ & $(-1,0,-1,0 )$  &  $(0,0,0,0)$  & $(-1,1,-1,1 )$  & $(0,0,0,0)$ \\
$B_2$ & $  (1,1,-s_{1},-s_{1} )$  &  $(0,0,0,0)$  & $(1,1,-1,-1 )$  & $(0,0,0,0)$ \\
$C$ &  $  (-1,-1,-1,-1 )$ & $(1,-1,1,-1 )$   & $s'(-1,-1,-1,-1)$  &  $ {\chi}_1(-1, 1,-1, 1) $  \\ 
\end{tabular}
\end{ruledtabular}
\end{table*}

\subsection{$\mathbb{C}\mathbb{P}^1$ action for (A$_1$)}\label{CP1Action}
In this section, we present the continuum-field theoretical description of the transition from the spin-liquid phase to altermagnetic order in the case of the A$_1$ phase. Following Read and Sachdev's procedure \cite{PhysRevLett.66.1773,Sachdev2012}, we parameterize the bosonic operators in A and B sublattices using slowly varying fields, i.e., 
\begin{align}
\label{parameterization1}
b_{A,\vec{r}_i,\alpha} &= {\psi}_{A,\alpha} (\vec{r}_{i}) e^{i \vec{Q} \cdot \vec{r}_{i}}, \\
b_{B,\vec{r}_i,\alpha} &= {\sigma}^y_{\alpha \beta}{\psi}_{B,\beta} (\vec{r}_{i}) e^{i \vec{Q} \cdot \vec{r}_{i}} \label{parameterization2}, 
\end{align}
where $\vec{Q}$ is the position in momentum space where spinon condensation takes place. With $\mathcal{A}_1<0$, in case of our ansatz for the A$_1$ phase (as described in \figref{fig:SchwingerBoson} of the main text), we found dispersion minimum to be at $\vec{Q}=(0, \pi)$. 
 To derive the continuum field theory, we plug equations (\ref{parameterization1}-\ref{parameterization2}) into the mean-field Hamiltonian (\ref{HMF1}), and carry out a gradient expansion, which leads to (in an action description)
\begin{equation}
\begin{split}
S &=\int \dfrac{d^2\vec{r}}{a^2} \int d\tau  \Big[ \big( {\psi}_{A,\alpha}^{\ast} \dfrac{d}{d \tau} {\psi}_{A,\alpha} + {\psi}_{B,\alpha}^{\ast} \dfrac{d}{d \tau} {\psi}_{B,\alpha} \big) - \mu \big( {\psi}_{A,\alpha}^{\ast} {\psi}_{A,\alpha} + {\psi}_{B,\alpha}^{\ast} {\psi}_{B,\alpha} \big) \\
 &+ 2\mathcal{A}_1 \big( {\psi}_{A,\alpha}{\psi}_{B,\alpha} + {\psi}_{A,\alpha}^{\ast} {\psi}_{B,\alpha}^{\ast} \big) - \dfrac{\mathcal{A}_1 a^2}{4} \big( \nabla {\psi}_{A,\alpha} \nabla {\psi}_{B,\alpha} + \nabla {\psi}_{A,\alpha}^{\ast} \nabla {\psi}_{B,\alpha}^{\ast} \big) \Big].
\end{split} 
\end{equation}
Next, we introduce two fields defined by 
$
{z}_{\alpha} = \dfrac{{\psi}_{A,\alpha}+{\psi}_{B,\alpha}^{\ast}}{\sqrt{2}}$ and $
{\pi}_{\alpha} = \dfrac{{\psi}_{A,\alpha}-{\psi}_{B,\alpha}^{\ast}}{\sqrt{2}} 
$
and rewrite the action as 
\begin{align}
\label{action1}
S =\int \dfrac{d^2\vec{r}}{a^2} \int d\tau  \Big[ \big( {z}_{\alpha}^{\ast} \dfrac{d}{d \tau} {\pi}_{\alpha} + {\pi}_{\alpha}^{\ast} \dfrac{d}{d \tau} {z}_{\alpha} \big)  + (2\mathcal{A}_1 - \mu)  |{z}_{\alpha}|^2  - (2\mathcal{A}_1+ \mu)  |{\pi}_{\alpha}|^2 -\dfrac{\mathcal{A}_1 a^2}{4} \big( | \nabla  {z}_{\alpha}|^2  - | \nabla {\pi}_{\alpha}|^2 \big)  \Big].
\end{align}
As can be seen from \equref{action1}, in this new form, ${\pi}_{\alpha}$ fields turn out to be massive, whereas $z_{\alpha}$ fields become massless near the critical ${\mu}_c = 2\mathcal{A}_1$, marking the phase transition from the U$(1)[0,0,0,\{0,0,0,0\},\{0,0,0,0\}]$ spin-liquid to altermagnetic order $A_{1}$. Note that the critical chemical potential ${\mu}_c $ is the same as the one obtained previously within the SB framework, exemplifying the consistency between both approaches. Finally, integrating the ${\pi}_{\alpha}$ fields and restoring gauge invariance by introduction of the gauge field $a_{\mu}$, we arrive at the following effective action 
\begin{equation}
S_{\text{eff}}= \int \dfrac{d^2\vec{r}}{4a} \int_0^{c \beta} d \bar{\tau} \Big[ |({\partial}_{\mu} -i a_{\mu})z_{\alpha}|^2 + \dfrac{{\Delta}^2}{c^2} |z_{\alpha}|^2 \Big].
\end{equation}
This essentially describes the massive spinons $z_{\alpha}$ coupled to a compact U(1) gauge field, where $c=|\mathcal{A}_1|a$ is the spinon velocity, ${\Delta}^2 = ({\mu}^2 - 4\mathcal{A}_1^2)$, and $a_{\bar{\tau}}=\dfrac{a_{\tau}}{c}$. 

\section{Electronic spectral function}\label{ElectronicSpcFun}
In this Appendix, we explain the details on how the electronic properties inside the fractionalized AM phases in \figref{fig:ElectronicSpectra} of the main text are computed. The total spin-fermion-model-like action, $\mathcal{S} = \mathcal{S}_e + \mathcal{S}_{c}+\mathcal{S}_{\Phi}$, we start from consists of three parts: first the non-interaction electronic action,
\begin{equation}
    \mathcal{S}_e = \int_0^\beta \diff \tau \left[ \sum_{\vec{R},\ell} c^\dagger_{\vec{R},\ell,\sigma} (\partial_\tau - \mu) c^\pdagger_{\vec{R},\ell,\sigma} - \sum_{\vec{R},\vec{R}',\ell,\ell'} t_{\vec{R},\ell;\vec{R}',\ell'} c^\dagger_{\vec{R},\ell,\sigma} c^\pdagger_{\vec{R}',\ell',\sigma} \right],
\end{equation}
where $c^\dagger_{\vec{R},\ell,\sigma}$ are the ($\tau$-dependent) electronic fields associated with the creation of an electron in the unit cell labeled by the Bravais lattice site $\vec{R}$, of sublattice $\ell=A,B$, and of spin $\sigma$. Note that $\mathcal{S}_e$ is invariant under global SR (summation over repeated $\sigma$ indices is implied). Although most of the following analysis is more generally valid, we later choose the hopping matrix elements $t_{\vec{R},\ell;\vec{R}',\ell'}$ as described in the main text with nearest and next-nearest neighbor hopping on the lattice in \figref{fig:classicalPD}(a). These electrons are coupling via [$\vec{\sigma} = (\sigma_x,\sigma_y,\sigma_z)^T$ is a vector of Pauli matrices]
\begin{equation}
    \mathcal{S}_c = \int_0^\beta \diff \tau \sum_{\vec{R},\ell} c^\dagger_{\vec{R},\ell,\sigma} \vec{\sigma}_{\sigma,\sigma'} c^\pdagger_{\vec{R},\ell,\sigma'} \cdot \vec{\Phi}_{\vec{R},\ell}
\end{equation}
to a bosonic field $\vec{\Phi}$ which describes fluctuations of the electronic spins. While at high-energies the action $\mathcal{S}_{\Phi}$ of $\vec{\Phi}$ can just be obtained via a Hubbard-Stratonovich decoupling of some bare electronic interactions, we here think of $\mathcal{S}$ as an effective low-energy action, just as in the celebrated spin-fermion model \cite{SpinFermionModel}, and take
\begin{equation}
    \mathcal{S}_{\Phi} = \frac{1}{4g_0}  \int_0^\beta \diff \tau \left[ \sum_{\vec{R},\ell} \left( \left(\partial_\tau \vec{\Phi}_{\vec{R},\ell}\right)^2  + V(\vec{\Phi}_{\vec{R},\ell}^2) \right) + \sum_{\vec{R},\vec{R}',\ell,\ell'} \mathcal{J}_{\vec{R},\ell;\vec{R}',\ell'} \vec{\Phi}_{\vec{R},\ell} \cdot \vec{\Phi}_{\vec{R}',\ell'}   \right]. \label{OriginalSPhi}
\end{equation}
Here $\mathcal{J}_{\vec{R},\ell;\vec{R}',\ell'}$ are effective exchange coupling constants, akin to $J_{j}$ in the spin model (\ref{eq:heisenbergmodel}).

To describe the fractionalized phases, characterized by topological order, we transform into a rotating reference frame in spin space via \cite{PhysRevLett.61.467,PhysRevLett.65.2462,SchriefferRotatingRefFrame}
\begin{equation}
    c_{\vec{R},\ell,\sigma}(\tau) = \sum_{\alpha=\pm} \left(R_{\vec{R},\ell}(\tau)\right)_{\sigma,\alpha}  \psi_{\vec{R},\ell,\alpha}(\tau). \label{RotatingReferenceFrame}
\end{equation}
In \equref{RotatingReferenceFrame}, $R$ are (space-time dependent, bosonic) SU(2) matrices, physically related to the spinons and, thus, to the Schwinger bosons in \equref{RewritingOfTheSpin} in the spin-liquid regime. While the spin is carried by $R$, the charge is governed by the fermionic ``chargon'' fields $\psi$; to indicate that the latter do not carry spin, we use $\alpha = \pm$ (rather than $\sigma=\uparrow,\downarrow$) to label its two components. As the physical electrons are invariant under the local reparameterization,
\begin{equation}
    R^\pdagger_{\vec{R},\ell}(\tau) \, \rightarrow \, R^\pdagger_{\vec{R},\ell}(\tau) V^\dagger_{\vec{R},\ell}(\tau), \quad \psi_{\vec{R},\ell}(\tau) \, \rightarrow \, V_{\vec{R},\ell}(\tau) \psi_{\vec{R},\ell}(\tau), \quad V^\dagger_{\vec{R},\ell}(\tau) V^\pdagger_{\vec{R},\ell}(\tau) = \mathbbm{1}, \label{SU2GaugeTrafo}
\end{equation}
inserting \equref{RotatingReferenceFrame} into $\mathcal{S}$, as we shall do in the following, will lead to an SU(2) gauge theory.

To discuss the action of spinons and chargons, we start with the coupling $\mathcal{S}_c$ which can be exactly rewritten as
\begin{equation}
    \mathcal{S}_c = \int_0^\beta \diff \tau \sum_{\vec{R},\ell} \psi^\dagger_{\vec{R},\ell,\alpha} \vec{\sigma}_{\alpha,\alpha'} \psi^\pdagger_{\vec{R},\ell,\alpha'} \cdot \vec{H}_{\vec{R},\ell}, \label{HiggsFieldCoupling}
\end{equation}
where $\vec{H}$ plays the role of a Higgs field, transforming under the adjoint representation of the SU(2) gauge transformation (\ref{SU2GaugeTrafo}). It is formally related to the collective boson $\vec{\Phi}$ via
\begin{equation}
    \vec{H}_{\vec{R},\ell} \cdot \vec{\sigma} = \vec{\Phi}_{\vec{R},\ell} \cdot R^\dagger_{\vec{R},\ell} \vec{\sigma} R^\pdagger_{\vec{R},\ell} \quad \text{and thus} \quad (\vec{\Phi}_{\vec{R},\ell})_a = \frac{1}{2} \text{tr}\left[\sigma_a R^\pdagger_{\vec{R},\ell} \vec{\sigma} R^\dagger_{\vec{R},\ell} \right] \cdot \vec{H}_{\vec{R},\ell} . \label{HiggPhiRelation}
\end{equation}
Inserting the parameterization (\ref{RotatingReferenceFrame}) into the non-interacting electronic part $\mathcal{S}_e$ of the action will lead to quartic terms; as in \refcite{OurPNAS}, we will decouple those into chargon and spinon parts. This leads to the chargon action
\begin{equation}
    \mathcal{S}_{\psi} = \int_0^\beta \diff \tau \left[  \sum_{\vec{R},\ell} \psi^\dagger_{\vec{R},\ell,\alpha} (\partial_\tau - \mu) \psi^\pdagger_{\vec{R},\ell,\alpha} - \sum_{\vec{R},\vec{R}',\ell,\ell'} t_{\vec{R},\ell;\vec{R}',\ell'} \psi^\dagger_{\vec{R},\ell,\alpha} (\mathcal{U}_{\vec{R},\ell;\vec{R}',\ell'})_{\alpha,\alpha'} \psi^\pdagger_{\vec{R}',\ell',\alpha'} \right], \label{ChargonActionGeneral}
\end{equation}
where $(\mathcal{U}_{\vec{R},\ell;\vec{R}',\ell'})_{\alpha,\alpha'} = \braket{(R^\dagger_{\vec{R},\ell} R^\pdagger_{\vec{R}',\ell'})_{\alpha,\alpha'}}$ and the spinon contribution
\begin{equation}
    \mathcal{S}_R^e = \int_0^\beta \diff \tau \, \text{tr} \left[ \sum_{\vec{R},\ell} \chi^T_{\vec{R},\ell;\vec{R},\ell} R^\dagger_{\vec{R},\ell} \partial_{\tau }R^\pdagger_{\vec{R},\ell} - \sum_{\vec{R},\vec{R}',\ell,\ell'} t_{\vec{R},\ell;\vec{R}',\ell'} \chi^T_{\vec{R},\ell;\vec{R}',\ell'} R^\dagger_{\vec{R},\ell} R^\pdagger_{\vec{R}',\ell'} \right],
\end{equation}
with the $2\times 2$ matrices $(\chi_{\vec{R},\ell;\vec{R}',\ell'})_{\alpha,\alpha'} = \braket{\psi^\dagger_{\vec{R},\ell,\alpha} \psi^\pdagger_{\vec{R}',\ell',\alpha'}}$.

Finally, plugging \equref{HiggPhiRelation} into $\mathcal{S}_\Phi$ in \equref{OriginalSPhi} leads to coupling terms between the Higgs field and the spinons. To describe the fractionalized phase with topological order, we follow previous works, see, e.g., \refcite{OurPNAS}, and keep the Higgs field at the saddle point value $\vec{H}_{\vec{R},\ell} \rightarrow \braket{\vec{H}_{\vec{R},\ell}}$. Note that $\braket{\vec{H}_{\vec{R},\ell}} \neq 0$ does \textit{not} break SR symmetry as the Higgs field is invariant under SRs [cf.~\equref{HiggsFieldCoupling}], as long as the bosons $R$ remain gapped/not condensed. We can now associate non-magnetic fractionalized states with each of the phases in the classical phase diagram in \figref{fig:classicalPD}(b) by choosing $\braket{\vec{H}_{\vec{R},\ell}}$ to be equal to the respective spin texture in the gauge where the bosons $R_{\vec{R},\ell}$ will condense in a spatially uniform way, $\braket{R_{\vec{R},\ell}} = R_0$, once the gap closes. For, in that case, we see in \equref{HiggPhiRelation} that $\braket{\vec{\Phi}_{\vec{R},\ell}}$ will be equal to $\braket{\vec{H}_{\vec{R},\ell}}$ apart from a global SR. To keep the discussion concise, we focus on phase A$_1$, the collinear AM, in the following; without loss of generality we choose a gauge where
\begin{equation}
    \braket{\vec{H}_{\vec{R},\ell}} = H_0 \hat{\vec{e}}_z s_\ell, \quad \text{with} \quad s_A = - s_B = 1.
\end{equation}
In this gauge, it is convenient to use the $\mathbb{C}\mathbb{P}^1$ parameterization 
\begin{equation}
    R_{\vec{R},\ell} = \begin{pmatrix} (z_{\vec{R},\ell})_\uparrow & -(z^*_{\vec{R},\ell})_\downarrow \\ (z_{\vec{R},\ell})_\downarrow & (z^*_{\vec{R},\ell})_\uparrow \end{pmatrix}, \quad \text{with} \quad z_{\vec{R},\ell}^\dagger z^\pdagger_{\vec{R},\ell} = 1. \label{RewritingRintofZ}
\end{equation}
The reason is that this leads to the intuitive form, $\vec{N}_{\vec{R},\ell} = z_{\vec{R},\ell}^\dagger \vec{\sigma} z^\pdagger_{\vec{R},\ell}$, of the normalized altermagnetic order parameter $\vec{N}_{\vec{R},\ell} = s_\ell \vec{\Phi}_{\vec{R},\ell}/|\vec{\Phi}_{\vec{R},\ell}|$ as readily follows from \equref{HiggPhiRelation}. We next express $\mathcal{S}_{\Phi}$ in \equref{OriginalSPhi} in terms of $\vec{N}_{\vec{R},\ell}$, yielding
\begin{equation}
    \mathcal{S}_{\Phi} = \frac{1}{4g}  \int_0^\beta \diff \tau \left[ \sum_{\vec{R},\ell} \left( \partial_\tau \vec{N}_{\vec{R},\ell}\right)^2   + \sum_{\vec{R},\vec{R}',\ell,\ell'} s_\ell s_{\ell'} \mathcal{J}_{\vec{R},\ell;\vec{R}',\ell'} \vec{N}_{\vec{R},\ell} \cdot \vec{N}_{\vec{R}',\ell'}   \right], \quad g=g_0/H_0^2,
\end{equation}
which we next study in the continuum limit; to this end, we introduce the continuum fields $\vec{n}(\tau,\vec{x})$ and $\vec{m}(\tau,\vec{x})$ with the constraints $\vec{n}^2=1$ and $\vec{n}\cdot \vec{m} = 0$ and write
\begin{equation}
    \vec{N}_{\vec{R},\ell}(\tau) = \vec{n}(\tau,\vec{R} + \vec{v}_\ell) \sqrt{1-\vec{m}^2(\tau,\vec{R}+\vec{v}_\ell)} + s_\ell \,\vec{m}(\tau,\vec{R}+\vec{v}_\ell), \quad\text{with}\quad \vec{v}_A = \hat{\vec{x}}/2,\,\, \vec{v}_B = \hat{\vec{y}}/2.
\end{equation}
Including nearest ($J_1$) and next-nearest-neighbor ($J_2$) exchange couplings, just like in the spin model we study, see \figref{fig:classicalPD}(a) of the main text, and expanding up to second order in gradients and $\vec{m}$, we find the non-linear sigma model
\begin{equation}
    \mathcal{S}_{\Phi} = \frac{1}{4g}  \int_0^\beta \diff \tau \diff^2\vec{x} \left[ \left( \partial_\tau \vec{n}\right)^2 + v^2 \left( \partial_x \vec{n}\right)^2  + v^2 \left( \partial_y \vec{n}\right)^2 + r \, \vec{m}^2  \right], \label{NLSM} 
\end{equation}
with $v^2 = J_1/2 -J_2/4$ and $r = 8 J_1$. As expected, the ferromagnetic fluctuations $\vec{m}$ are massive, $r>0$, and can thus be neglected, i.e., $\vec{N}_{\vec{R},\ell}(\tau) \approx \vec{n}(\tau,\vec{R} + \vec{v}_\ell)$. The AM fluctuations $\vec{n}$ are governed by the usual relativistic form in \equref{NLSM}, which in turn can be restated \cite{PhysRevB.81.115129,PhysRevB.42.4568} as the $\mathbb{C}\mathbb{P}^1$ model 
\begin{equation}
    \mathcal{S}_{\Phi}^{z,c} = g^{-1}\int \diff \tau \diff^2\vec{x} \,| (\partial_\mu - ia_\mu) z_{\alpha} |^2, \label{CP1ContAction}
\end{equation}
where $a_\mu$ is an emergent U(1) gauge field; this is exactly the same continuum field theory we derived for the SB mean-field ansatz for phase A$_1$ in \appref{CP1Action}. As in \refcite{OurPNAS}, we will neglect the gauge-field fluctuations, which are not expected to change the results qualitatively \cite{PhysRevB.75.235122}, and work on the lattice, i.e., use
\begin{equation}
    \mathcal{S}_{\Phi}^{z,c} = g^{-1} T\sum_{\Omega_n}\sum_{\vec{q}} z^\dagger_{n,\vec{q}} z^\pdagger_{n,\vec{q}} \left[ \Omega_n^2 + \mathcal{E}^2_{\vec{q}} \right], \qquad \mathcal{E}^2_{\vec{q}} = 2J^2(2-\cos q_x - \cos q_y) + \Delta^2 \label{SpinonActionWeUse}
\end{equation}
for the bosons in \equref{RewritingRintofZ}; here $\Delta$ is the bosonic gap that has to be finite in the spin-liquid regime with topological order and is determined by the condition $\braket{z_{\vec{R},\ell}^\dagger z^\pdagger_{\vec{R},\ell}} = 1$. For the numerics presented in \figref{fig:ElectronicSpectra}(d-f) of the main text, we used $J/t_1=1$ and $\Delta/t_1=0.01$.  

With the spinon action (\ref{SpinonActionWeUse}) at hand, we can simplify the form of the chargon action in \equref{ChargonActionGeneral}. In the gauge leading to \equref{SpinonActionWeUse}, we immediate conclude that $\mathcal{U}_{\vec{R},\ell;\vec{R}',\ell'} = \text{diag} (\braket{z^\dagger_{\vec{R},\ell}z^\pdagger_{\vec{R}',\ell'}},\braket{z^\dagger_{\vec{R}',\ell'}z^\pdagger_{\vec{R},\ell}})$. As the spinon dispersion $\mathcal{E}_{\vec{q}}$ is even in $\vec{q}$, the two terms on the diagonal of $\mathcal{U}_{\vec{R},\ell;\vec{R}',\ell'}$ are identical (and real) such that we are left with
\begin{equation}
    \mathcal{U}_{\vec{R},\ell;\vec{R}',\ell'} = \mathbbm{1} Z_{\vec{R},\ell;\vec{R}',\ell'}, \qquad Z_{\vec{R},\ell;\vec{R}',\ell'} \in \mathbb{R}.
\end{equation}
As the spinon (and chargon) theory is further invariant under all spatial symmetries, there will only be two independent values, $Z$ and $Z'$, renormalizing, respectively, the nearest- ($t$) and next-nearest-neighbor hopping ($t'$) of the chargons relative to the bare electrons, $t \rightarrow Z t$ and $t'\rightarrow Z't'$. One can determine $Z$ and $Z'$ by solving the chargon and spinon theory self-consistently \cite{OurPNAS}, yielding order-$1$ values. To facilitate the comparison of the non-interacting spectra and of the spectral function in the fractionalized theory in \figref{fig:ElectronicSpectra} of the main text, we simply set $Z=Z'=1$ in the following. Including the contribution from the Higgs field, the full chargon action can be written as 
\begin{equation}
    \mathcal{S}_\psi + \mathcal{S}_c = T \sum_{\omega_n} \frac{1}{N} \sum_{\vec{k}} \psi^\dagger_{\vec{k},\alpha} \left[-i \omega_n + \epsilon_{\vec{k}} \tau_0 + \vec{g}^\alpha_{\vec{k}} \cdot \vec{\tau} \right] \psi^\pdagger_{\vec{k},\alpha},
\end{equation}
where $\tau_j$ are Pauli matrices in sublattice space and
\begin{align}
    \epsilon_{\vec{k}} = -t'\left( \cos k_x + \cos k_y \right)-\mu,& \, (\vec{g}^\alpha_{\vec{k}})_x + i (\vec{g}^\alpha_{\vec{k}})_y = t \left( 1 + e^{ik_x} + e^{i(k_x-k_y)} + e^{-ik_y}\right),\\ \, (\vec{g}^\alpha_{\vec{k}})_z &= -t'\left( \cos k_x - \cos k_y \right) + H_0 \alpha.
\end{align}
Consequently, the chargon Green's function (a matrix in sublattice space) reads as
\begin{equation}
    G^\psi_{\alpha,\alpha'}(i\omega_n,\vec{k}) = \delta_{\alpha,\alpha'}\frac{i \omega_n - \epsilon_{\vec{k}}+\vec{g}^\alpha_{\vec{k}} \cdot \vec{\tau}}{\left(i\omega_n - E^+_{\vec{k},\alpha}\right)\left(i\omega_n - E^-_{\vec{k},\alpha}\right)}, \quad E^\pm_{\vec{k},\alpha} = \epsilon_{\vec{k}} \pm |\vec{g}^\alpha_{\vec{k}}|.
\end{equation}

Finally, we can compute the electron Green's function, $G_e(i\omega_n,\vec{k})$, which by virtue of the rewriting (\ref{RotatingReferenceFrame}) becomes a convolution of the chargon and spinon Green's function, see also the diagram in \figref{fig:ElectronicSpectra}(c). After straightforward algebra, we find
\begin{equation}
    \left[G_e(i\omega_n,\vec{k})\right]_{\ell,\sigma;\ell',\sigma'} = \delta_{\sigma,\sigma'}T \sum_{\Omega_m} \frac{1}{N} \sum_{\vec{q}} \sum_{\alpha=\pm} G_z(i\Omega_m,\vec{q}) G^\psi_{\alpha,\alpha}(i\omega_n-i\Omega_m,\vec{k}-\vec{q}) e^{i\vec{q}(\vec{v}_{\ell'}-\vec{v}_{\ell})},
\end{equation}
where $G_z(i\Omega_n,\vec{q}) = g/(\Omega_n^2 + \mathcal{E}^2_{\vec{q}})$ is the spinon Green's function associated with \equref{SpinonActionWeUse}. We can directly see that the convolution with the spinons restored full SU(2) SR invariance ($\propto \delta_{\sigma,\sigma'}$). More explicitly, the resultant spectral function we plot in the main text is written as
\begin{align}
    \mathcal{A}_{\vec{k}}(\omega) &= -\frac{1}{4\pi}\text{Im}\,\text{tr}\left[G_e(\omega + i \eta,\vec{k})\right] \\ &= -\frac{1}{\pi N} \text{Im} \sum_{\vec{q}} \left( T \sum_{\Omega_m} \sum_{\alpha=\pm} \frac{g}{\Omega_m^2 + \mathcal{E}^2_{\vec{q}}} \frac{i \omega_n-i\Omega_m - \epsilon_{\vec{k}-\vec{q}}+\vec{g}^\alpha_{\vec{k}-\vec{q}} \cdot \vec{\tau}}{\left(i\omega_n-i\Omega_m  - E^+_{\vec{k}-\vec{q},\alpha}\right)\left(i\omega_n-i\Omega_m - E^-_{\vec{k}-\vec{q},\alpha}\right)} \right)_{i\omega_n \rightarrow \omega + i \eta}.
\end{align}
As the spinon gap $\Delta$ has a more direct physical meaning than the coupling constant $g$, we fix $\Delta$ and then think of $g$ as a function of $\Delta$ (and not vice versa), $g=g(\Delta)$. It was shown in \refcite{OurPNAS} that the above approximations leave the frequency sum rule,
\begin{equation}
    \int_{-\infty}^\infty \diff \omega \, A_{\vec{k}}(\omega) = 1, \label{FreqSumRule}
\end{equation}
untouched. We use \equref{FreqSumRule} to determine $g$ for each $\Delta$, yielding the non-trivial check that a \textit{single} $g$ can ensure that \equref{FreqSumRule} holds at \textit{all} $\vec{k}$.

\end{appendix}

\end{document}